\documentclass[%
 reprint,
superscriptaddress, %uncommented
 amsmath,amssymb,
 aps,
]{revtex4-2}
\usepackage{physics}
\usepackage{graphicx}% Include figure files
\usepackage{dcolumn}% Align table columns on decimal point
\usepackage{bm}% bold math
\usepackage[colorlinks, citecolor=blue]{hyperref}% add hypertext capabilities
\usepackage{xcolor} %for color in text
\newcommand{\ceil}[1]{\left\lceil #1 \right\rceil} %define ceil

\begin{document}

%Sensing photon addition to nonclassical light using the Wasserstein distance %title (a)

\title{Optimal sensing of photon addition and subtraction on nonclassical light}
\author{Soumyabrata Paul}
\email{soumyabrata@physics.iitm.ac.in}
\affiliation{Department of Physics, Indian Institute of Technology Madras, Chennai 600036, India}
\affiliation{Center for Quantum Information, Communication and Computing (CQuICC), Indian Institute of Technology Madras, Chennai 600036, India}
\author{Arman}
\affiliation{Department of Physical Sciences, Indian Institute of Science Education and Research Kolkata, Mohanpur 741246, India}
\author{S. Lakshmibala}
\affiliation{Center for Quantum Information, Communication and Computing (CQuICC), Indian Institute of Technology Madras, Chennai 600036, India}
\author{Prasanta K. Panigrahi}
\affiliation{Department of Physical Sciences, Indian Institute of Science Education and Research Kolkata, Mohanpur 741246, India}
\affiliation{Center for Quantum Science and Technology (CQST), Siksha `O' Anusandhan University, Khandigiri Square, Bhubaneswar 751030, India}
\author{S. Ramanan}
\affiliation{Department of Physics, Indian Institute of Technology Madras, Chennai 600036, India}
\affiliation{Center for Quantum Information, Communication and Computing (CQuICC), Indian Institute of Technology Madras, Chennai 600036, India}
\author{V. Balakrishnan}
\affiliation{Center for Quantum Information, Communication and Computing (CQuICC), Indian Institute of Technology Madras, Chennai 600036, India}

\date{\today}% It is always \today, today,
             %  but any date may be explicitly specified

\begin{abstract}
We demonstrate that the Wasserstein distance $W_{1}$ corresponding to optical 
tomograms of nonclassical states faithfully captures changes that arise due to 
photon addition to, or subtraction from, these states.
$W_{1}$ is a true measure of distance in the quantum state space, and is sensitive to 
the underlying interference structures that arise in the tomogram after changes in 
the photon number.
Our procedure is universally applicable to the cat and squeezed states, the former 
displaying the characteristic negativity in its Wigner function, while the latter 
does not do so.
We explicate this in the case of the squeezed vacuum and even coherent states and 
show that photon addition (or subtraction) is mirrored in the shift in the 
intensity of specific regions in the tomogram. Further, we examine the 
dependence of $W_{1}$ on the squeezing parameter, and its sensitivity to different 
quadratures.
\end{abstract}

%\keywords{Suggested keywords}%Use showkeys class option if keyword
                              %display desired
\maketitle

%\tableofcontents

\section{Introduction\label{sec:introduction}}

An important aspect of quantum metrology is to increase the extent of sensing of nonclassical properties of quantum states beyond the standard quantum limit~\cite{Giovannetti:2011}. The strongly squeezed vacuum state sent through a Mach-Zehnder interferometer~\cite{Baune:2015}, the two-mode squeezed vacuum state~\cite{Anisimov:2010},  and the photon added two-mode squeezed vacuum state~\cite{Ouyang:2016}  are examples of ideal candidates for this purpose. 
It has been shown that photon addition to the squeezed vacuum state (SVS) and the even coherent state (ECS) leads to a shift in the phase space interference features in the corresponding Wigner function as well as modifications in the associated sub-Planck structures, leading to significant metrological changes~\cite{Akhtar:2023, Panigrahi:2021}.
However, state reconstruction is in general an arduous task for continuous variable systems with large dimensional Hilbert spaces. In this paper we describe an optimal program for identifying photon number changes in nonclassical light solely from its tomogram. The tomogram is the starting point in any state reconstruction procedure.
While, in principle, the inverse Radon transformation of the tomogram gives the Wigner function, the stepwise procedure to be implemented for this purpose is, in general, challenging, particularly for multipartite systems.
As a first step, we demonstrate the efficacy of our procedure for both the SVS and the ECS. The details concerning the ECS are discussed in Sec.~IV of the supplementary material (SM).
The procedure that we outline would possibly be optimal for multipartite states as well.

The SVS and other squeezed states of light are of great current interest, as they are very useful in beating the standard quantum limit~\cite{Feng:2024}, in quantum error correction~\cite{Qin:2024, Schlegel:2022}, enhancing quantum sensing~\cite{Zhang:2024}, in the metrology of absorption and gain parameters~\cite{Kamble:2024, Li:2021}, generation of non-standard types of photon blockade~\cite{Kowalewska:2019} and provide the potential for optical switching between the normal and superradiant phases of light~\cite{Zhu:2020}. 
The SVS has diverse applications in, for instance, photoelectric detection~\cite{Schnabel:2016}, gravitational wave detectors~\cite{Caves:1981, LIGO:2013}, and parameter estimation in bilinear Hamiltonians~\cite{Paris:2009}. 
It is therefore not surprising that optimal methods to generate single-mode squeezed light~\cite{Andersen:2016, Park:2024}, and to obtain a quick and reliable estimation of its extent of degradation owing to decoherence, using machine learning tools~\cite{Hsien-Yi:2022} are being examined extensively.

Investigations on the effect of the addition of a photon to, or its removal from the light field have been undertaken for decades (see, for instance,~\cite{Stefano:2003, Parigi:2007, Biswas:2007, Grangier:2007, Masahiro:2008, Xu:2012, Cerf:2012, Das:2016, Barnett:2018, Takase:2021, Grebien:2022, Hongbin:2023, Tomoda:2024, Fadrny:2024}).
The performance of quantum operations by addition or removal of photons from traveling light beams has been investigated~\cite{Kim:2008, Fiurasek:2009}.
It has been proposed~\cite{Zhang:1992} that if atoms are subjected to appropriate nonlinear processes induced in a cavity which is initialized with an SVS, photon addition to the SVS could be possible in the short time dynamics.
This is analogous to photon addition to coherent states~\cite{Agarwal:1991}.
Both photon addition to and subtraction from multimode and entangled optical fields~\cite{Thapliyal:2024}, and generation of highly nonclassical states by the addition of photons to twin beams of light~\cite{Perina:2024}, have been achieved experimentally.

In general, changes in the photon number of an SVS lead to increased nonclassicality of the resultant state and introduce non-Gaussian characteristics~\cite{Walschaers:2021, Biagi:2022}. 
It is known in metrology that in detecting phase space shifts, the SVS provides better global sensitivity when compared to the cat and Fock states. This sensitivity is further enhanced by performing non-Gaussian operations such as photon addition to the squeezed state~\cite{Gorecki:2022}.
This has diverse applications in quantum computation~\cite{Lloyd:1999, Menicucci:2006, Bartlett:2002, Mari:2012}.
Boson sampling procedures to investigate quantum supremacy use photon added and subtracted SVS~\cite{Olson:2015}.

Different aspects pertaining to photon addition to nonclassical light~\cite{Zhang:1992, Quesne:2001, Xu:2012, Yuan:2019, Bohloul:2024} have been reported in the literature. Photon addition is important for enhanced sensing. For instance, the sensitivity of intensity detection in an ${\rm SU}(1,1)$ interferometer can approach the Heisenberg limit if photons are added to the SVS in the input port~\cite{Guo:2018}. Quick generation of heralded optical cat states by addition of photons has been experimentally realised~\cite{Chen:2024}.

Photon subtracted squeezed vacuum states have been experimentally produced in different ways. For instance, these states have been obtained from a continuous-wave optical parametric amplifier~\cite{Neergaard-Nielsen:2006}, and by using periodically-poled KTiOPO$_{4}$ crystal as a nonlinear medium~\cite{Wakui:2007}.  One and two photon subtracted SVS have been generated at telecommunication wavelengths, and identified using a titanium superconducting transition edge sensor~\cite{Namekata:2010}. Subtraction of up to three photons has been shown to result in large-amplitude coherent state superpositions~\cite{Gerrits:2010}.

In what follows we point out (among other results) a simple and elegant procedure to detect photon addition to, or subtraction from the SVS, by observing and quantifying changes in its tomogram.
This procedure is also applicable to other nonclassical states, and is more viable than performing state reconstruction.

An important feature of the tomogram of a quantum state is that it is a set of probability distributions in different quadratures, obeying the laws of classical probability theory. 
In general, we shall therefore compare the new state (after photon addition or subtraction) with the reference state, quantifying the difference,  using the notion of a distance between probability distributions. In what follows we use the Wasserstein distance $W_{1}$~\cite{Vaserstein:1969} for our purpose. We point out that $W_{1}$ satisfies the properties of a distance metric in contrast to other quantifiers such as the Kullback-Leibler divergence~\cite{KL:1951} and the Bhattacharyya distance~\cite{Bhattacharyya:1943}, which are useful in defining entanglement indicators~\cite{Sharmila:2019}.

Each tomogram is represented as an image or pattern as explained later. (Hence $W_{1}$ is a pattern measure and has extensive applications in image processing and machine learning protocols.) In earlier literature $W_{1}$ has been effectively used to quantify the divergence between different eigenstates of the simple harmonic oscillator, a particle in a one-dimensional box and the state of light at different instants when it interacts with an atomic medium~\cite{Soumyabrata:2024}. 
Here, we demonstrate that quantifying the changes in the tomogram using $W_{1}$ suffices to identify changes in the photon number. The 
significant advantage (as mentioned earlier) is that state reconstruction can be avoided for this purpose. This exercise also reveals interesting links between the SVS and the ECS from a tomographic point of view, for it augments prior knowledge on `Janus-faced' partners reported in the literature~\cite{Shanta:1994, Laha:2018} (see  Sec.~IV of SM).

This paper is organized as follows. In Sec.~\ref{sec:tomographic_pattens_and_wasserstein_distance} we discuss the role played by $W_{1}$ in comparing tomographic patterns. In Sec.~\ref{sec:photon_added_svs} we use $W_{1}$ to distinguish between the SVS and the photon added SVS. A similar exercise is carried out in Sec.~\ref{sec:photon_subtracted_svs} on the effect of photon subtraction from the SVS. We conclude with a brief summary and outlook in Sec.~\ref{sec:concluding_remarks}.

\section{Tomographic patterns and the Wasserstein distance\label{sec:tomographic_pattens_and_wasserstein_distance}}

Consider a single-mode radiation field with photon creation and annihilation operators $\hat{a}^{\dagger}$ and $\hat{a}$ respectively. The set of rotated quadrature operators \cite{Ibort:2009},
\begin{equation}
\mathbb{\hat{X}_{\theta}} = \left( \hat{a}^{\dagger} e^{i \theta} + \hat{a} e^{-i \theta} \right) / \sqrt{2}.
\label{eq:quadop_singlemode}
\end{equation}
Here $\theta$ ($0 \leqslant \theta < \pi$) is the phase of the local oscillator in the standard homodyne measurement setup. The $x$-quadrature (resp. $p$-quadrature) corresponds to $\theta = 0$ (resp. $\theta = \pi/2$). Equation ~\eqref{eq:quadop_singlemode} constitutes a quorum of observables which carry complete information about the single-mode state with density matrix $\hat{\rho}$. The corresponding optical tomogram $w(X_{\theta}, \theta)$ \cite{Lvovsky:2009} is defined as
\begin{equation} 
w(X_{\theta}, \theta) = \langle X_{\theta}, \theta | \hat{\rho} | X_{\theta}, \theta \rangle,
\label{eq:w_singlemode}
\end{equation}
where 
\begin{equation}
\mathbb{\hat{X}_{\theta}} |X_{\theta}, \theta \rangle = X_{\theta}|X_{\theta}, \theta \rangle.
\label{eq:quadop_eigenval_eq}
\end{equation}
For a normalized pure state $|\psi\rangle$ it is evident that the tomogram is given by $|\psi(X_{\theta}, \theta)|^{2}$ for a given $\theta$. The experimentally measured quantity is $w(X_{\theta}, \theta)$ (i.e., only the diagonal elements of ${\hat \rho}$ for a given $\theta$). The completeness relation is given by
\begin{equation} 
\int_{-\infty}^{\infty} dX_{\theta}~w(X_{\theta}, \theta) = 1 ~ \forall ~ \theta,
\label{eq:w_singlemode_normalization}
\end{equation}
where $\{|X_{\theta}, \theta\rangle\}$ forms a complete basis for every $\theta$. For computational purposes it is advantageous to expand $w(X_{\theta}, \theta)$ in the Fock basis~\cite{Filippov:2011}.

The tomogram exhibits the symmetry property
\begin{equation} 
w(X_{\theta}, \theta + \pi) = w(-X_{\theta}, \theta),
\label{eq:w_singlemode_symmetry}
\end{equation}
and is commonly represented as a pattern with $X_{\theta}$ as the abscissa, and $\theta$ as the ordinate. Single-mode optical tomograms for the SVS and some of its photon added and subtracted counterparts are shown in Fig.~\ref{fig:fig_tomogram_SVS_PASVS_PSSVS_DPASVS_TPASVS_DPSSVS_TPSSVS_r_1OverSqrt2_phi_0_panel}. For state reconstruction, although it is sufficient to work with the range $0 \leqslant \theta < \pi$, the tomogram plotted for $0 \leqslant \theta < 2\pi$ helps visualize various features better.

In what follows, we will use the Wasserstein distance to compare two tomograms $w_{{\rm A}}(X_{\theta_{{\rm A}}}, \theta_{{\rm A}})$ and $w_{{\rm B}}(X_{\theta_{{\rm B}}}, \theta_{{\rm B}})$ corresponding to two distinct single-mode radiation fields A and B. Without loss of generality (for our purposes), we first set $\theta_{{\rm A}} = \theta_{{\rm B}} = 0$ (the $x$-quadrature for each field). The two normalized one-dimensional probability density functions (PDFs) are represented by $f(x)$ and $g(x)$, corresponding to the fields A and B, respectively. The Wasserstein distance $W_{1}$ essentially quantifies the minimum cost in transporting one PDF to the other. It is given in terms of the corresponding cumulative distribution functions (CDFs) $F(x) = \int_{-\infty}^{x} dy~f(y)$ and $G(x) = \int_{-\infty}^{x} dy~g(y)$ as
\begin{equation}
W_{1}(F, G) = \int_{-\infty}^{\infty} dx \left| F(x) - G(x) \right|.
\label{eq:wasserstein_distance}
\end{equation}
We note that $W_{1}$ is a true distance metric, in the sense that $W_{1}(F, F) = 0$, $W_{1}(F, G) = W_{1}(G, F)$, and it satisfies the triangle inequality. We now proceed to obtain $W_{1}$ between the SVS and its photon added and photon subtracted counterparts. Table~1 in SM lists the states and their relevant probability amplitudes.

\vspace{2.5ex} %FORCE A VERTICAL SPACE BY HAND, REMOVE IF NOT REQUIRED
\section{Photon Added Squeezed Vacuum States\label{sec:photon_added_svs}}
%%%%%%%%%%%%%%%%%%%%%%%%%%%%%%%%%%%%%%%%%%%%%%%%%%%%
%%%%%%%%%%%%%%%%%%%%%%%%%%%%%%%%%%%%%%%%%%%%%%%%%%%%
\begin{figure}[t]
    \centering
    \includegraphics[width=0.45\textwidth]{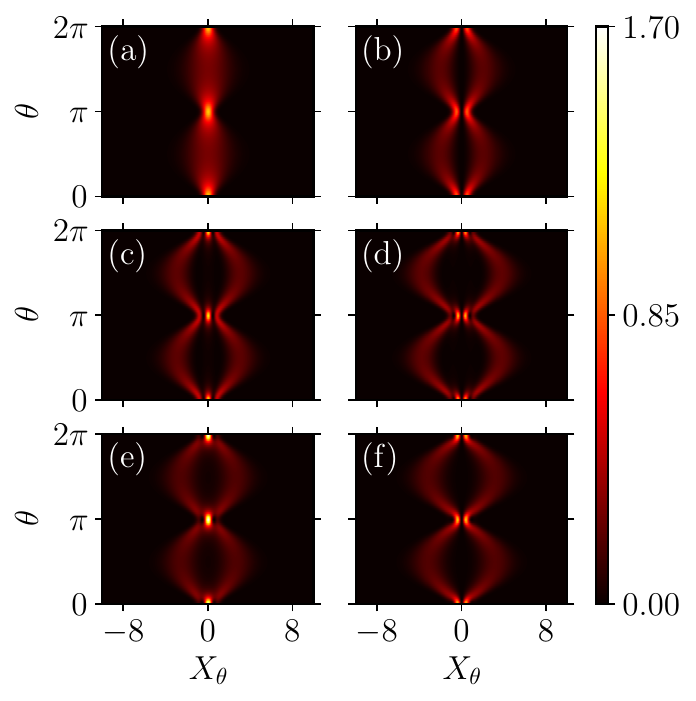}
    \caption{Left to right: Single-mode optical tomograms corresponding to (a) $|\xi\rangle$, (b) $|\xi,1\rangle$ (or $|\xi,-1\rangle$), (c) $|\xi,2\rangle$, (d) $|\xi,3\rangle$, (e) $|\xi,-2\rangle$ and (f) $|\xi,-3\rangle$. The squeezing parameter $\xi = re^{i\phi}$ ($r = 1/\sqrt{2}$ and $\phi = 0$). The bright central region at $\theta=0$, $\pi$, $2\pi$ in (a) indicates a peak in the PDF. Addition or subtraction of a single photon to the SVS $|\xi\rangle$ destroys this peak (the central dark region in (b)). With the addition or subtraction of two photons the bright central region reappears ((c) and (e)). With the addition or subtraction of three photons the bright central region disappears, and is replaced by a central dark region ((d) and (f)).}
    \label{fig:fig_tomogram_SVS_PASVS_PSSVS_DPASVS_TPASVS_DPSSVS_TPSSVS_r_1OverSqrt2_phi_0_panel}
\end{figure}
%%%%%%%%%%%%%%%%%%%%%%%%%%%%%%%%%%%%%%%%%%%%%%%%%%%%
%%%%%%%%%%%%%%%%%%%%%%%%%%%%%%%%%%%%%%%%%%%%%%%%%%%%
The squeezed vacuum $|\xi\rangle={\hat S}(\xi)|0\rangle$, where the squeezing operator ${\hat S}(\xi) = \exp \left[ \frac{1}{2} ({\xi}^{*} {\hat a}^2 - \xi {\hat a}^{\dagger 2}) \right]$. Here $\xi = r e^{i\phi}$, $r \geqslant 0$ and $\phi \in [0, 2\pi)$. For $\phi = 0$ (resp. $\phi = \pi$) squeezing is along the $x$-quadrature (resp. $p$-quadrature). It can be shown that~\cite{Fisher:1984}
\begin{widetext}
\begin{equation}
{\hat S}(\xi) = \exp\left[-\frac{1}{2}e^{-i\phi}(\tanh r)~{\hat a}^{\dagger 2} \right] \left[ (\cosh r)^{-1/2}\sum_{n=0}^{\infty}\frac{({\rm sech}~r - 1)^{n}}{n!} (\hat{a}^{\dagger}\hat{a})^{n} \right] \exp\left[ \frac{1}{2}e^{i\phi}(\tanh r)~{\hat a}^{2} \right].
\label{eq:squeezing_op_disentangled}
\end{equation}
\end{widetext}
Hence $|\xi\rangle$ is obtained by the action of an exponential operator which is a function of $a^{\dagger 2}$ alone on $|0\rangle$, i.e., $\exp(g {\hat a}^{\dagger 2})|0\rangle$, where $g$ is a scalar. We will draw attention to this feature in Sec.~IV of SM, where we discuss the ECS.

The SVS $|\xi\rangle$ can be expressed in the Fock basis as
\begin{equation}
|\xi\rangle = \frac{1}{\sqrt{\cosh r}} \sum_{n=0}^{\infty} \frac{\sqrt{(2n)}!}{2^n n!} e^{i n \phi} (- \tanh r)^{n}~|2n\rangle.
\label{eq:xi_svs_fock_expansion}
\end{equation}
By normalizing $\hat{a}^{\dagger m}|\xi\rangle$ ($m = 1, 2, \dots$) we obtain the $m$-photon added SVS $|\xi,m\rangle$. The expansion of $|\xi,m\rangle$ in the Fock basis is given by~\cite{Zhang:1992, Quesne:2001}
\begin{equation}
|\xi,m\rangle = \frac{1}{{\mathcal{N}^{1/2}_{m}}} \sum_{n=0}^{\infty} \frac{\sqrt{(2n + m)}!}{2^n n!} e^{i n \phi} (- \tanh r)^{n}~ |2n+m\rangle,
\label{eq:xi_m_photon_added_svs}
\end{equation}
with $\mathcal{N}_{m} = m! (\cosh r)^{m+1} P_{m}(\cosh r)$, where $P_{m}$ is the Legendre polynomial of degree $m$.
The tomogram for the SVS does not display alternating intensity fringes. However, on addition of photons to the SVS, such fringes appear. Depending on the number of photons added there are shifts in the alternating bright and dark intensity fringes seen at $\theta = 0$, $\pi$ and $2\pi$, with $\phi = 0$ (see Fig.~\ref{fig:fig_tomogram_SVS_PASVS_PSSVS_DPASVS_TPASVS_DPSSVS_TPSSVS_r_1OverSqrt2_phi_0_panel} here, and its zoomed-in version in Fig.~1 of SM).
The number of dark bands corresponds to the number of added photons (Figs.~\ref{fig:fig_tomogram_SVS_PASVS_PSSVS_DPASVS_TPASVS_DPSSVS_TPSSVS_r_1OverSqrt2_phi_0_panel} (b), (c) and (d)).

%%%%%%%%%%%%%%%%%%%%%%%%%%%%%%%%%%%%%%%%%%%%%%%%%%%%
%%%%%%%%%%%%%%%%%%%%%%%%%%%%%%%%%%%%%%%%%%%%%%%%%%%%
\begin{figure}[h]
    \centering
    \includegraphics[width=0.45\textwidth]{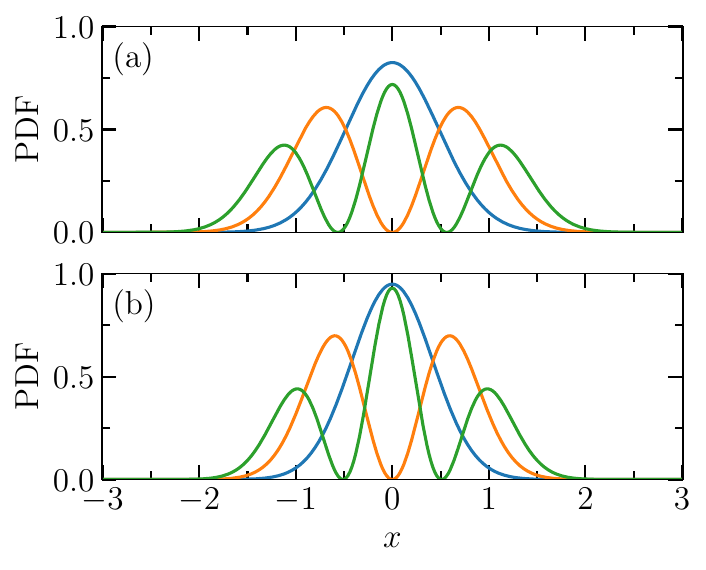}
    \caption{Probability distribution function (PDF) along the $x$-quadrature corresponding to $|\xi\rangle$ (blue), $|\xi,1\rangle$ (orange), and $|\xi,2\rangle$ (green) for $r = $ (a) $0.38$ and (b) $0.52$. The other parameter values are as in Fig.~\ref{fig:fig_tomogram_SVS_PASVS_PSSVS_DPASVS_TPASVS_DPSSVS_TPSSVS_r_1OverSqrt2_phi_0_panel}. At $x = 0$, the PDF for the SVS $|\xi\rangle$ and the two-photon added SVS $|\xi,2\rangle$ are maximum, while the one-photon added SVS $|\xi,1\rangle$ has a corresponding minimum. This is reflected in the shift in the intensity patterns seen in Fig.~\ref{fig:fig_tomogram_SVS_PASVS_PSSVS_DPASVS_TPASVS_DPSSVS_TPSSVS_r_1OverSqrt2_phi_0_panel}.}
    \label{fig:PDF_SVS_panel}
\end{figure}
%%%%%%%%%%%%%%%%%%%%%%%%%%%%%%%%%%%%%%%%%%%%%%%%%%%%
%%%%%%%%%%%%%%%%%%%%%%%%%%%%%%%%%%%%%%%%%%%%%%%%%%%%

\begin{figure}[h]
    \centering
    \includegraphics[width=0.45\textwidth]{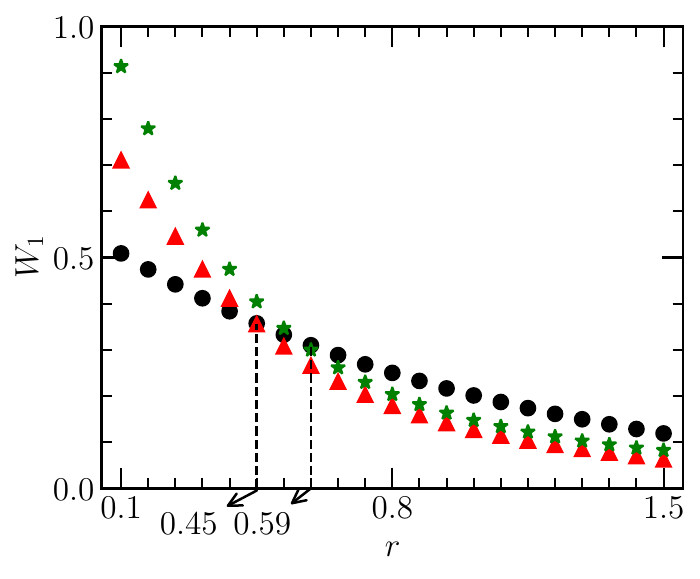}
    \caption{Wasserstein distance $W_{1}$ between $|\xi\rangle$ and $|\xi, 1\rangle$ [or $|\xi,-1\rangle$] (black circles), $|\xi\rangle$ and $|\xi, 2\rangle$ (red triangles), and $|\xi\rangle$ and $|\xi, 3\rangle$ (green asterisks) as a function of the squeezing parameter $r$, along the $x$-quadrature. The other parameters are as in Fig.~\ref{fig:fig_tomogram_SVS_PASVS_PSSVS_DPASVS_TPASVS_DPSSVS_TPSSVS_r_1OverSqrt2_phi_0_panel}. The crossover between $W_1$ for one photon addition and two photon addition is at $r \approx 0.45$ and that between one photon addition (the same as one photon subtraction) and three photon addition is at $r \approx 0.59$. To identify the extent of photon addition and to distinguish between photon added counterparts using $W_{1}$ it is therefore important to avoid the neighborhood of values of $r$ where crossovers occur.}
    \label{fig:WD_PhAddedSVS}
\end{figure}

To quantify the extent of the shift in the fringes seen in the tomograms with 
the addition of photons, we now examine the corresponding PDFs along the $x$-quadrature 
(see Fig.~\ref{fig:PDF_SVS_panel}). The corresponding CDFs are calculated
from the PDFs. We compute $W_{1}$ as a function of the squeezing parameter $r$ (Fig.~\ref{fig:WD_PhAddedSVS}) for
a range of experimentally relevant values of $r$. We treat the SVS $|\xi\rangle$ as the reference state.
In what follows, $W_{1}(|\xi\rangle, |\xi, m\rangle)$ refers to the Wasserstein distance computed between $|\xi\rangle$ and $|\xi,m\rangle$, in a specific quadrature, in this case the $x$-quadrature.
First consider the two plots, namely, $W_{1}(|\xi\rangle, |\xi,1\rangle)$ (black circles),
and $W_{1}(|\xi\rangle, |\xi,2\rangle)$ (red triangles).
These two distances are equal at a crossover value of $r \approx 0.45$.
Similarly, there is a crossover at $r \approx 0.59$ between the plots for $W_{1}(|\xi\rangle, |\xi,1\rangle)$ (black circles),
and $W_{1}(|\xi\rangle, |\xi,3\rangle)$ (green asterisks), as seen in Fig.~\ref{fig:WD_PhAddedSVS}.
Such crossovers can be seen in some other quadratures as well as shown in Fig.~2 of
SM. (We have also given examples of quadratures where such crossovers are absent.)
To identify the extent of photon addition and to distinguish between photon 
added counterparts using $W_{1}$, it is therefore important to avoid the values of $r$ where 
crossovers occur.

The quadrature variance
along the direction of squeezing computed for the SVS and the photon added SVS are 
shown in Fig.~5 of SM. The exponent in the fall-off as a function of the squeezing
parameter in the case of one and two photon addition to the SVS is different from that of the
SVS as expected. 

As mentioned in the Introduction, the SVS and the ECS are related in an interesting manner
to each other. It is therefore useful to summarize 
briefly here the results obtained for the photon added ECS, in SM.
The tomograms themselves have interference patterns in the $p$-quadrature in contrast to the
SVS. However, as in the case of the SVS, intensity shifts are seen on addition of photons,
with a bright (resp. dark) central fringe on addition of an even (resp. odd) number of
photons. In contrast to the SVS, a crossover is seen in the $p$-quadrature ($\theta=\pi/2$) and its neighborhood.

%%%%%%%%%%%%%%%%%%%%%%%%%%%%%%%%%%%%%%%%%%%%%%%%%%%%
%%%%%%%%%%%%%%%%%%%%%%%%%%%%%%%%%%%%%%%%%%%%%%%%%%%%
%%%%%%%%%%%%%%%%%%%%%%%%%%%%%%%%%%%%%%%%%%%%%%%%%%%%
%%%%%%%%%%%%%%%%%%%%%%%%%%%%%%%%%%%%%%%%%%%%%%%%%%%%

\section{Photon Subtracted Squeezed Vacuum States\label{sec:photon_subtracted_svs}}
Photon subtraction from the SVS is carried out in theory by repeated application of $\hat{a}$ on the SVS and normalizing the resulting 
state. The $m$-photon subtracted SVS is given by
\begin{equation}
|\xi, -m\rangle = \frac{1}{{\mathcal{K}^{1/2}_{-m}}} \sum_{n=\ceil{m/2}}^{\infty} \frac{(2n)! e^{i n \phi} (-\tanh r)^{n}}{2^n n! \sqrt{(2n-m)!}} ~ |2n-m\rangle,
\label{eq:xi_m_photon_subtracted_svs}
\end{equation}
where
\begin{equation}
\mathcal{K}_{-m} = \sum_{n=\ceil{m/2}}^{\infty}\left[\frac{(2n)!}{2^{n}n!}\right]^{2}\frac{(\tanh r)^{2n}}{(2n-m)!}
\label{eq:normalization_m_photon_subtracted_svs}
\end{equation}
and $\ceil{m/2}$ denotes the smallest integer $\geqslant m/2$.
It is clear from Eqs.~\eqref{eq:xi_m_photon_added_svs} and~\eqref{eq:xi_m_photon_subtracted_svs} that one photon addition or subtraction operations on $|\xi\rangle$, are equivalent~\cite{Biswas:2007}.
Tomograms corresponding to photon subtraction from the SVS are shown in Figs.~\ref{fig:fig_tomogram_SVS_PASVS_PSSVS_DPASVS_TPASVS_DPSSVS_TPSSVS_r_1OverSqrt2_phi_0_panel} (b), (e) and (f) above. Their zoomed-in versions are given in SM (Figs.~1 (b), (e) and (f)).

Analogous to Fig.~\ref{fig:WD_PhAddedSVS}, we have also examined the crossovers in the plots of the Wasserstein distances versus $r$ between $|\xi\rangle$ and the different photon subtracted states $|\xi,-m\rangle$ ($m=1,2,3$). 
In the cases we have considered, there is no crossover along the $x$-quadrature.
There are however, other quadratures, where such crossovers occur, as seen in Fig.~7 of SM.
In both photon addition and subtraction, it is evident that the value of $r$ at a crossover
depends on both the quadrature and the states considered.
Hence, the neighborhood of crossovers should be avoided in order to see a clear distinction between states using the Wasserstein distance.

\section{Concluding remarks\label{sec:concluding_remarks}}

In continuous variable systems, quantum states could suffer from errors arising through photon loss or gain. A precise  estimate of the number of such photons is a prerequisite for error correction. High precision detection is possible using states with large sensitivity or small variance in those quadratures where the changes arise due to such errors \cite{Preskill:2001}. We have shown that, for appropriate ranges of the state parameter, the effect of photon addition or subtraction can be captured in the Wasserstein distances computed between the nonclassical reference state (the squeezed vacuum, for instance) and the corresponding photon added/subtracted states. We have further shown that similar results hold in the case of photon addition to the even cat state over an appropriate range of parameter values. It is worth noting that noise estimation gets enhanced by addition of photons to the original nonclassical state \cite{Akhtar:2023, Arman:2024}.

Our approach is based on examining the qualitative features of the relevant optical tomograms (histograms of experimental data). In particular we have examined the shifts in the intensities of nonclassical light in specific quadratures, arising due to photon addition or subtraction.  The PDFs comprising the relevant tomograms corroborate our findings.
We have demonstrated  the usefulness of this tomographic approach both for identifying qualitative changes and for quantifying such changes 
using the Wasserstein distance. In particular, we have identified specific small ranges of values of state parameters (crossover values) which are not 
sufficiently sensitive to photon addition to, or removal from the nonclassical reference state. We have also identified a sufficiently wide range of experimentally accessible parameter values where it is possible to quantify, with precision, the extent of change in the photon number by computing the Wasserstein distance directly from the tomograms.

In conclusion, we have presented an experimentally viable, tomogram-based measure of quantum state discrimination (the Wasserstein distance) for sensing subtle differences between 
nonclassical states. The Wasserstein distance is a true metrological measure, that discriminates between states that arise from the addition and 
subtraction of photons. Its advantages are illustrated through examples of nonclassical states, ranging from the squeezed vacuum to cat 
states and their generalizations. Our approach suggests the possibility of sensing changes in the photon number directly from the optical 
tomogram, circumventing detailed state reconstruction. This could be potentially useful for identifying such changes in generic multimode states where reconstructing the Wigner function from the tomogram could pose considerable challenges.

\begin{acknowledgments}
We acknowledge partial support through funds from Mphasis to the Centre for Quantum Information, Communication and Computing (CQuICC), Indian Institute of Technology Madras. SL and VB thank the Department of Physics, Indian Institute of Technology Madras for infrastructural support. We thank the referees for their valuable suggestions.
\end{acknowledgments}

%\appendix
%\section{Appendixes}
%\section{A little more on appendixes}
%\subsection{\label{app:subsec}A subsection in an appendix}

\bibliography{references}% Produces the bibliography via BibTeX.

%apsrev4-2.bst 2019-01-14 (MD) hand-edited version of apsrev4-1.bst
%Control: key (0)
%Control: author (8) initials jnrlst
%Control: editor formatted (1) identically to author
%Control: production of article title (0) allowed
%Control: page (0) single
%Control: year (1) truncated
%Control: production of eprint (0) enabled
\begin{thebibliography}{71}%
\makeatletter
\providecommand \@ifxundefined [1]{%
 \@ifx{#1\undefined}
}%
\providecommand \@ifnum [1]{%
 \ifnum #1\expandafter \@firstoftwo
 \else \expandafter \@secondoftwo
 \fi
}%
\providecommand \@ifx [1]{%
 \ifx #1\expandafter \@firstoftwo
 \else \expandafter \@secondoftwo
 \fi
}%
\providecommand \natexlab [1]{#1}%
\providecommand \enquote  [1]{``#1''}%
\providecommand \bibnamefont  [1]{#1}%
\providecommand \bibfnamefont [1]{#1}%
\providecommand \citenamefont [1]{#1}%
\providecommand \href@noop [0]{\@secondoftwo}%
\providecommand \href [0]{\begingroup \@sanitize@url \@href}%
\providecommand \@href[1]{\@@startlink{#1}\@@href}%
\providecommand \@@href[1]{\endgroup#1\@@endlink}%
\providecommand \@sanitize@url [0]{\catcode `\\12\catcode `\$12\catcode `\&12\catcode `\#12\catcode `\^12\catcode `\_12\catcode `\%12\relax}%
\providecommand \@@startlink[1]{}%
\providecommand \@@endlink[0]{}%
\providecommand \url  [0]{\begingroup\@sanitize@url \@url }%
\providecommand \@url [1]{\endgroup\@href {#1}{\urlprefix }}%
\providecommand \urlprefix  [0]{URL }%
\providecommand \Eprint [0]{\href }%
\providecommand \doibase [0]{https://doi.org/}%
\providecommand \selectlanguage [0]{\@gobble}%
\providecommand \bibinfo  [0]{\@secondoftwo}%
\providecommand \bibfield  [0]{\@secondoftwo}%
\providecommand \translation [1]{[#1]}%
\providecommand \BibitemOpen [0]{}%
\providecommand \bibitemStop [0]{}%
\providecommand \bibitemNoStop [0]{.\EOS\space}%
\providecommand \EOS [0]{\spacefactor3000\relax}%
\providecommand \BibitemShut  [1]{\csname bibitem#1\endcsname}%
\let\auto@bib@innerbib\@empty
%</preamble>
\bibitem [{\citenamefont {Giovannetti}\ \emph {et~al.}(2011)\citenamefont {Giovannetti}, \citenamefont {Lloyd},\ and\ \citenamefont {Maccone}}]{Giovannetti:2011}%
  \BibitemOpen
  \bibfield  {author} {\bibinfo {author} {\bibfnamefont {V.}~\bibnamefont {Giovannetti}}, \bibinfo {author} {\bibfnamefont {S.}~\bibnamefont {Lloyd}},\ and\ \bibinfo {author} {\bibfnamefont {L.}~\bibnamefont {Maccone}},\ }\bibfield  {title} {\bibinfo {title} {Advances in quantum metrology},\ }\href {https://doi.org/10.1038/nphoton.2011.35} {\bibfield  {journal} {\bibinfo  {journal} {Nature Photonics}\ }\textbf {\bibinfo {volume} {5}},\ \bibinfo {pages} {222} (\bibinfo {year} {2011})}\BibitemShut {NoStop}%
\bibitem [{\citenamefont {Baune}\ \emph {et~al.}(2015)\citenamefont {Baune}, \citenamefont {Gniesmer}, \citenamefont {Sch\"{o}nbeck}, \citenamefont {Vollmer}, \citenamefont {Fiur\'{a}\v{s}ek},\ and\ \citenamefont {Schnabel}}]{Baune:2015}%
  \BibitemOpen
  \bibfield  {author} {\bibinfo {author} {\bibfnamefont {C.}~\bibnamefont {Baune}}, \bibinfo {author} {\bibfnamefont {J.}~\bibnamefont {Gniesmer}}, \bibinfo {author} {\bibfnamefont {A.}~\bibnamefont {Sch\"{o}nbeck}}, \bibinfo {author} {\bibfnamefont {C.~E.}\ \bibnamefont {Vollmer}}, \bibinfo {author} {\bibfnamefont {J.}~\bibnamefont {Fiur\'{a}\v{s}ek}},\ and\ \bibinfo {author} {\bibfnamefont {R.}~\bibnamefont {Schnabel}},\ }\bibfield  {title} {\bibinfo {title} {Strongly squeezed states at 532 nm based on frequency up-conversion},\ }\href {https://doi.org/10.1364/OE.23.016035} {\bibfield  {journal} {\bibinfo  {journal} {Opt. Express}\ }\textbf {\bibinfo {volume} {23}},\ \bibinfo {pages} {16035} (\bibinfo {year} {2015})}\BibitemShut {NoStop}%
\bibitem [{\citenamefont {Anisimov}\ \emph {et~al.}(2010)\citenamefont {Anisimov}, \citenamefont {Raterman}, \citenamefont {Chiruvelli}, \citenamefont {Plick}, \citenamefont {Huver}, \citenamefont {Lee},\ and\ \citenamefont {Dowling}}]{Anisimov:2010}%
  \BibitemOpen
  \bibfield  {author} {\bibinfo {author} {\bibfnamefont {P.~M.}\ \bibnamefont {Anisimov}}, \bibinfo {author} {\bibfnamefont {G.~M.}\ \bibnamefont {Raterman}}, \bibinfo {author} {\bibfnamefont {A.}~\bibnamefont {Chiruvelli}}, \bibinfo {author} {\bibfnamefont {W.~N.}\ \bibnamefont {Plick}}, \bibinfo {author} {\bibfnamefont {S.~D.}\ \bibnamefont {Huver}}, \bibinfo {author} {\bibfnamefont {H.}~\bibnamefont {Lee}},\ and\ \bibinfo {author} {\bibfnamefont {J.~P.}\ \bibnamefont {Dowling}},\ }\bibfield  {title} {\bibinfo {title} {Quantum metrology with two-mode squeezed vacuum: Parity detection beats the {H}eisenberg limit},\ }\href {https://doi.org/10.1103/PhysRevLett.104.103602} {\bibfield  {journal} {\bibinfo  {journal} {Phys. Rev. Lett.}\ }\textbf {\bibinfo {volume} {104}},\ \bibinfo {pages} {103602} (\bibinfo {year} {2010})}\BibitemShut {NoStop}%
\bibitem [{\citenamefont {Ouyang}\ \emph {et~al.}(2016)\citenamefont {Ouyang}, \citenamefont {Wang},\ and\ \citenamefont {Zhang}}]{Ouyang:2016}%
  \BibitemOpen
  \bibfield  {author} {\bibinfo {author} {\bibfnamefont {Y.}~\bibnamefont {Ouyang}}, \bibinfo {author} {\bibfnamefont {S.}~\bibnamefont {Wang}},\ and\ \bibinfo {author} {\bibfnamefont {L.}~\bibnamefont {Zhang}},\ }\bibfield  {title} {\bibinfo {title} {Quantum optical interferometry via the photon-added two-mode squeezed vacuum states},\ }\href {https://doi.org/10.1364/JOSAB.33.001373} {\bibfield  {journal} {\bibinfo  {journal} {J. Opt. Soc. Am. B}\ }\textbf {\bibinfo {volume} {33}},\ \bibinfo {pages} {1373} (\bibinfo {year} {2016})}\BibitemShut {NoStop}%
\bibitem [{\citenamefont {Akhtar}\ \emph {et~al.}(2023)\citenamefont {Akhtar}, \citenamefont {Wu}, \citenamefont {Peng}, \citenamefont {Liu},\ and\ \citenamefont {Xianlong}}]{Akhtar:2023}%
  \BibitemOpen
  \bibfield  {author} {\bibinfo {author} {\bibfnamefont {N.}~\bibnamefont {Akhtar}}, \bibinfo {author} {\bibfnamefont {J.}~\bibnamefont {Wu}}, \bibinfo {author} {\bibfnamefont {J.-X.}\ \bibnamefont {Peng}}, \bibinfo {author} {\bibfnamefont {W.-M.}\ \bibnamefont {Liu}},\ and\ \bibinfo {author} {\bibfnamefont {G.}~\bibnamefont {Xianlong}},\ }\bibfield  {title} {\bibinfo {title} {Sub-{P}lanck structures and sensitivity of the superposed photon-added or photon-subtracted squeezed-vacuum states},\ }\href {https://doi.org/10.1103/PhysRevA.107.052614} {\bibfield  {journal} {\bibinfo  {journal} {Phys. Rev. A}\ }\textbf {\bibinfo {volume} {107}},\ \bibinfo {pages} {052614} (\bibinfo {year} {2023})}\BibitemShut {NoStop}%
\bibitem [{\citenamefont {Arman}\ \emph {et~al.}(2021)\citenamefont {Arman}, \citenamefont {Tyagi},\ and\ \citenamefont {Panigrahi}}]{Panigrahi:2021}%
  \BibitemOpen
  \bibfield  {author} {\bibinfo {author} {\bibnamefont {Arman}}, \bibinfo {author} {\bibfnamefont {G.}~\bibnamefont {Tyagi}},\ and\ \bibinfo {author} {\bibfnamefont {P.~K.}\ \bibnamefont {Panigrahi}},\ }\bibfield  {title} {\bibinfo {title} {Photon added cat state: phase space structure and statistics},\ }\href {https://doi.org/10.1364/OL.415713} {\bibfield  {journal} {\bibinfo  {journal} {Opt. Lett.}\ }\textbf {\bibinfo {volume} {46}},\ \bibinfo {pages} {1177} (\bibinfo {year} {2021})}\BibitemShut {NoStop}%
\bibitem [{\citenamefont {Feng}\ \emph {et~al.}(2024)\citenamefont {Feng}, \citenamefont {Zhang},\ and\ \citenamefont {Wei}}]{Feng:2024}%
  \BibitemOpen
  \bibfield  {author} {\bibinfo {author} {\bibfnamefont {X.~N.}\ \bibnamefont {Feng}}, \bibinfo {author} {\bibfnamefont {M.}~\bibnamefont {Zhang}},\ and\ \bibinfo {author} {\bibfnamefont {L.~F.}\ \bibnamefont {Wei}},\ }\bibfield  {title} {\bibinfo {title} {Beating the standard quantum limit electronic field sensing by simultaneously using quantum entanglement and squeezing},\ }\href {https://doi.org/10.1103/PhysRevLett.132.220801} {\bibfield  {journal} {\bibinfo  {journal} {Phys. Rev. Lett.}\ }\textbf {\bibinfo {volume} {132}},\ \bibinfo {pages} {220801} (\bibinfo {year} {2024})}\BibitemShut {NoStop}%
\bibitem [{\citenamefont {Qin}\ \emph {et~al.}(2024)\citenamefont {Qin}, \citenamefont {Miranowicz},\ and\ \citenamefont {Nori}}]{Qin:2024}%
  \BibitemOpen
  \bibfield  {author} {\bibinfo {author} {\bibfnamefont {W.}~\bibnamefont {Qin}}, \bibinfo {author} {\bibfnamefont {A.}~\bibnamefont {Miranowicz}},\ and\ \bibinfo {author} {\bibfnamefont {F.}~\bibnamefont {Nori}},\ }\bibfield  {title} {\bibinfo {title} {Exponentially improved dispersive qubit readout with squeezed light},\ }\bibfield  {journal} {\bibinfo  {journal} {arXiv:2402.12044 [quant-ph]}\ }\href {https://doi.org/10.48550/arXiv.2402.12044} {10.48550/arXiv.2402.12044} (\bibinfo {year} {2024})\BibitemShut {NoStop}%
\bibitem [{\citenamefont {Schlegel}\ \emph {et~al.}(2022)\citenamefont {Schlegel}, \citenamefont {Minganti},\ and\ \citenamefont {Savona}}]{Schlegel:2022}%
  \BibitemOpen
  \bibfield  {author} {\bibinfo {author} {\bibfnamefont {D.~S.}\ \bibnamefont {Schlegel}}, \bibinfo {author} {\bibfnamefont {F.}~\bibnamefont {Minganti}},\ and\ \bibinfo {author} {\bibfnamefont {V.}~\bibnamefont {Savona}},\ }\bibfield  {title} {\bibinfo {title} {Quantum error correction using squeezed {S}chr\"odinger cat states},\ }\href {https://doi.org/10.1103/PhysRevA.106.022431} {\bibfield  {journal} {\bibinfo  {journal} {Phys. Rev. A}\ }\textbf {\bibinfo {volume} {106}},\ \bibinfo {pages} {022431} (\bibinfo {year} {2022})}\BibitemShut {NoStop}%
\bibitem [{\citenamefont {Zhang}\ \emph {et~al.}(2024)\citenamefont {Zhang}, \citenamefont {Wang}, \citenamefont {Zhang}, \citenamefont {Jiao}, \citenamefont {Zuo}, \citenamefont {\c{S}ahin K.~\"{O}zdemir}, \citenamefont {Qiu}, \citenamefont {Nori},\ and\ \citenamefont {Jing}}]{Zhang:2024}%
  \BibitemOpen
  \bibfield  {author} {\bibinfo {author} {\bibfnamefont {S.-D.}\ \bibnamefont {Zhang}}, \bibinfo {author} {\bibfnamefont {J.}~\bibnamefont {Wang}}, \bibinfo {author} {\bibfnamefont {Q.}~\bibnamefont {Zhang}}, \bibinfo {author} {\bibfnamefont {Y.-F.}\ \bibnamefont {Jiao}}, \bibinfo {author} {\bibfnamefont {Y.-L.}\ \bibnamefont {Zuo}}, \bibinfo {author} {\bibnamefont {\c{S}ahin K.~\"{O}zdemir}}, \bibinfo {author} {\bibfnamefont {C.-W.}\ \bibnamefont {Qiu}}, \bibinfo {author} {\bibfnamefont {F.}~\bibnamefont {Nori}},\ and\ \bibinfo {author} {\bibfnamefont {H.}~\bibnamefont {Jing}},\ }\bibfield  {title} {\bibinfo {title} {Squeezing-enhanced quantum sensing with quadratic optomechanics},\ }\href {https://doi.org/10.1364/OPTICAQ.523480} {\bibfield  {journal} {\bibinfo  {journal} {Optica Quantum}\ }\textbf {\bibinfo {volume} {2}},\ \bibinfo {pages} {222} (\bibinfo {year} {2024})},\ \bibinfo {note} {and references therein}\BibitemShut {NoStop}%
\bibitem [{\citenamefont {Kamble}\ \emph {et~al.}(2024)\citenamefont {Kamble}, \citenamefont {Wang},\ and\ \citenamefont {Agarwal}}]{Kamble:2024}%
  \BibitemOpen
  \bibfield  {author} {\bibinfo {author} {\bibfnamefont {M.}~\bibnamefont {Kamble}}, \bibinfo {author} {\bibfnamefont {J.}~\bibnamefont {Wang}},\ and\ \bibinfo {author} {\bibfnamefont {G.~S.}\ \bibnamefont {Agarwal}},\ }\bibfield  {title} {\bibinfo {title} {Quantum metrology of absorption and gain parameters using two-mode bright squeezed light},\ }\href {https://doi.org/10.1103/PhysRevA.109.053715} {\bibfield  {journal} {\bibinfo  {journal} {Phys. Rev. A}\ }\textbf {\bibinfo {volume} {109}},\ \bibinfo {pages} {053715} (\bibinfo {year} {2024})}\BibitemShut {NoStop}%
\bibitem [{\citenamefont {Li}\ \emph {et~al.}(2021)\citenamefont {Li}, \citenamefont {Li}, \citenamefont {Scully},\ and\ \citenamefont {Agarwal}}]{Li:2021}%
  \BibitemOpen
  \bibfield  {author} {\bibinfo {author} {\bibfnamefont {F.}~\bibnamefont {Li}}, \bibinfo {author} {\bibfnamefont {T.}~\bibnamefont {Li}}, \bibinfo {author} {\bibfnamefont {M.~O.}\ \bibnamefont {Scully}},\ and\ \bibinfo {author} {\bibfnamefont {G.~S.}\ \bibnamefont {Agarwal}},\ }\bibfield  {title} {\bibinfo {title} {Quantum advantage with seeded squeezed light for absorption measurement},\ }\href {https://doi.org/10.1103/PhysRevApplied.15.044030} {\bibfield  {journal} {\bibinfo  {journal} {Phys. Rev. Appl.}\ }\textbf {\bibinfo {volume} {15}},\ \bibinfo {pages} {044030} (\bibinfo {year} {2021})},\ \bibinfo {note} {and references therein}\BibitemShut {NoStop}%
\bibitem [{\citenamefont {Kowalewska-Kud\l{}aszyk}\ \emph {et~al.}(2019)\citenamefont {Kowalewska-Kud\l{}aszyk}, \citenamefont {Abo}, \citenamefont {Chimczak}, \citenamefont {Pe\ifmmode~\check{r}\else \v{r}\fi{}ina}, \citenamefont {Nori},\ and\ \citenamefont {Miranowicz}}]{Kowalewska:2019}%
  \BibitemOpen
  \bibfield  {author} {\bibinfo {author} {\bibfnamefont {A.}~\bibnamefont {Kowalewska-Kud\l{}aszyk}}, \bibinfo {author} {\bibfnamefont {S.~I.}\ \bibnamefont {Abo}}, \bibinfo {author} {\bibfnamefont {G.}~\bibnamefont {Chimczak}}, \bibinfo {author} {\bibfnamefont {J.}~\bibnamefont {Pe\ifmmode~\check{r}\else \v{r}\fi{}ina}}, \bibinfo {author} {\bibfnamefont {F.}~\bibnamefont {Nori}},\ and\ \bibinfo {author} {\bibfnamefont {A.}~\bibnamefont {Miranowicz}},\ }\bibfield  {title} {\bibinfo {title} {Two-photon blockade and photon-induced tunneling generated by squeezing},\ }\href {https://doi.org/10.1103/PhysRevA.100.053857} {\bibfield  {journal} {\bibinfo  {journal} {Phys. Rev. A}\ }\textbf {\bibinfo {volume} {100}},\ \bibinfo {pages} {053857} (\bibinfo {year} {2019})}\BibitemShut {NoStop}%
\bibitem [{\citenamefont {Zhu}\ \emph {et~al.}(2020)\citenamefont {Zhu}, \citenamefont {Ping}, \citenamefont {Yang},\ and\ \citenamefont {Agarwal}}]{Zhu:2020}%
  \BibitemOpen
  \bibfield  {author} {\bibinfo {author} {\bibfnamefont {C.~J.}\ \bibnamefont {Zhu}}, \bibinfo {author} {\bibfnamefont {L.~L.}\ \bibnamefont {Ping}}, \bibinfo {author} {\bibfnamefont {Y.~P.}\ \bibnamefont {Yang}},\ and\ \bibinfo {author} {\bibfnamefont {G.~S.}\ \bibnamefont {Agarwal}},\ }\bibfield  {title} {\bibinfo {title} {Squeezed light induced symmetry breaking superradiant phase transition},\ }\href {https://doi.org/10.1103/PhysRevLett.124.073602} {\bibfield  {journal} {\bibinfo  {journal} {Phys. Rev. Lett.}\ }\textbf {\bibinfo {volume} {124}},\ \bibinfo {pages} {073602} (\bibinfo {year} {2020})}\BibitemShut {NoStop}%
\bibitem [{\citenamefont {Vahlbruch}\ \emph {et~al.}(2016)\citenamefont {Vahlbruch}, \citenamefont {Mehmet}, \citenamefont {Danzmann},\ and\ \citenamefont {Schnabel}}]{Schnabel:2016}%
  \BibitemOpen
  \bibfield  {author} {\bibinfo {author} {\bibfnamefont {H.}~\bibnamefont {Vahlbruch}}, \bibinfo {author} {\bibfnamefont {M.}~\bibnamefont {Mehmet}}, \bibinfo {author} {\bibfnamefont {K.}~\bibnamefont {Danzmann}},\ and\ \bibinfo {author} {\bibfnamefont {R.}~\bibnamefont {Schnabel}},\ }\bibfield  {title} {\bibinfo {title} {Detection of 15 d{B} squeezed states of light and their application for the absolute calibration of photoelectric quantum efficiency},\ }\href {https://doi.org/10.1103/PhysRevLett.117.110801} {\bibfield  {journal} {\bibinfo  {journal} {Phys. Rev. Lett.}\ }\textbf {\bibinfo {volume} {117}},\ \bibinfo {pages} {110801} (\bibinfo {year} {2016})}\BibitemShut {NoStop}%
\bibitem [{\citenamefont {Caves}(1981)}]{Caves:1981}%
  \BibitemOpen
  \bibfield  {author} {\bibinfo {author} {\bibfnamefont {C.~M.}\ \bibnamefont {Caves}},\ }\bibfield  {title} {\bibinfo {title} {Quantum-mechanical noise in an interferometer},\ }\href {https://doi.org/10.1103/PhysRevD.23.1693} {\bibfield  {journal} {\bibinfo  {journal} {Phys. Rev. D}\ }\textbf {\bibinfo {volume} {23}},\ \bibinfo {pages} {1693} (\bibinfo {year} {1981})}\BibitemShut {NoStop}%
\bibitem [{\citenamefont {{The LIGO Scientific Collaboration}}(2013)}]{LIGO:2013}%
  \BibitemOpen
  \bibfield  {author} {\bibinfo {author} {\bibnamefont {{The LIGO Scientific Collaboration}}},\ }\bibfield  {title} {\bibinfo {title} {Enhanced sensitivity of the {LIGO} gravitational wave detector by using squeezed states of light},\ }\href {https://doi.org/10.1038/nphoton.2013.177} {\bibfield  {journal} {\bibinfo  {journal} {Nature Photonics}\ }\textbf {\bibinfo {volume} {7}},\ \bibinfo {pages} {613} (\bibinfo {year} {2013})}\BibitemShut {NoStop}%
\bibitem [{\citenamefont {Gaiba}\ and\ \citenamefont {Paris}(2009)}]{Paris:2009}%
  \BibitemOpen
  \bibfield  {author} {\bibinfo {author} {\bibfnamefont {R.}~\bibnamefont {Gaiba}}\ and\ \bibinfo {author} {\bibfnamefont {M.~G.}\ \bibnamefont {Paris}},\ }\bibfield  {title} {\bibinfo {title} {Squeezed vacuum as a universal quantum probe},\ }\href {https://doi.org/https://doi.org/10.1016/j.physleta.2009.01.026} {\bibfield  {journal} {\bibinfo  {journal} {Physics Letters A}\ }\textbf {\bibinfo {volume} {373}},\ \bibinfo {pages} {934} (\bibinfo {year} {2009})}\BibitemShut {NoStop}%
\bibitem [{\citenamefont {Andersen}\ \emph {et~al.}(2016)\citenamefont {Andersen}, \citenamefont {Gehring}, \citenamefont {Marquardt},\ and\ \citenamefont {Leuchs}}]{Andersen:2016}%
  \BibitemOpen
  \bibfield  {author} {\bibinfo {author} {\bibfnamefont {U.~L.}\ \bibnamefont {Andersen}}, \bibinfo {author} {\bibfnamefont {T.}~\bibnamefont {Gehring}}, \bibinfo {author} {\bibfnamefont {C.}~\bibnamefont {Marquardt}},\ and\ \bibinfo {author} {\bibfnamefont {G.}~\bibnamefont {Leuchs}},\ }\bibfield  {title} {\bibinfo {title} {30 years of squeezed light generation},\ }\href {https://doi.org/10.1088/0031-8949/91/5/053001} {\bibfield  {journal} {\bibinfo  {journal} {Physica Scripta}\ }\textbf {\bibinfo {volume} {91}},\ \bibinfo {pages} {053001} (\bibinfo {year} {2016})}\BibitemShut {NoStop}%
\bibitem [{\citenamefont {Park}\ \emph {et~al.}(2024)\citenamefont {Park}, \citenamefont {Stokowski}, \citenamefont {Ansari}, \citenamefont {Gyger}, \citenamefont {Multani}, \citenamefont {Celik}, \citenamefont {Hwang}, \citenamefont {Dean}, \citenamefont {Mayor}, \citenamefont {McKenna}, \citenamefont {Fejer},\ and\ \citenamefont {Safavi-Naeini}}]{Park:2024}%
  \BibitemOpen
  \bibfield  {author} {\bibinfo {author} {\bibfnamefont {T.}~\bibnamefont {Park}}, \bibinfo {author} {\bibfnamefont {H.}~\bibnamefont {Stokowski}}, \bibinfo {author} {\bibfnamefont {V.}~\bibnamefont {Ansari}}, \bibinfo {author} {\bibfnamefont {S.}~\bibnamefont {Gyger}}, \bibinfo {author} {\bibfnamefont {K.~K.~S.}\ \bibnamefont {Multani}}, \bibinfo {author} {\bibfnamefont {O.~T.}\ \bibnamefont {Celik}}, \bibinfo {author} {\bibfnamefont {A.~Y.}\ \bibnamefont {Hwang}}, \bibinfo {author} {\bibfnamefont {D.~J.}\ \bibnamefont {Dean}}, \bibinfo {author} {\bibfnamefont {F.}~\bibnamefont {Mayor}}, \bibinfo {author} {\bibfnamefont {T.~P.}\ \bibnamefont {McKenna}}, \bibinfo {author} {\bibfnamefont {M.~M.}\ \bibnamefont {Fejer}},\ and\ \bibinfo {author} {\bibfnamefont {A.}~\bibnamefont {Safavi-Naeini}},\ }\bibfield  {title} {\bibinfo {title} {Single-mode squeezed-light generation and tomography with an integrated optical parametric oscillator},\ }\href {https://doi.org/10.1126/sciadv.adl1814} {\bibfield  {journal}
  {\bibinfo  {journal} {Science Advances}\ }\textbf {\bibinfo {volume} {10}},\ \bibinfo {pages} {eadl1814} (\bibinfo {year} {2024})}\BibitemShut {NoStop}%
\bibitem [{\citenamefont {Hsieh}\ \emph {et~al.}(2022)\citenamefont {Hsieh}, \citenamefont {Chen}, \citenamefont {Wu}, \citenamefont {Chen}, \citenamefont {Ning}, \citenamefont {Huang}, \citenamefont {Wu},\ and\ \citenamefont {Lee}}]{Hsien-Yi:2022}%
  \BibitemOpen
  \bibfield  {author} {\bibinfo {author} {\bibfnamefont {H.-Y.}\ \bibnamefont {Hsieh}}, \bibinfo {author} {\bibfnamefont {Y.-R.}\ \bibnamefont {Chen}}, \bibinfo {author} {\bibfnamefont {H.-C.}\ \bibnamefont {Wu}}, \bibinfo {author} {\bibfnamefont {H.~L.}\ \bibnamefont {Chen}}, \bibinfo {author} {\bibfnamefont {J.}~\bibnamefont {Ning}}, \bibinfo {author} {\bibfnamefont {Y.-C.}\ \bibnamefont {Huang}}, \bibinfo {author} {\bibfnamefont {C.-M.}\ \bibnamefont {Wu}},\ and\ \bibinfo {author} {\bibfnamefont {R.-K.}\ \bibnamefont {Lee}},\ }\bibfield  {title} {\bibinfo {title} {Extract the degradation information in squeezed states with machine learning},\ }\href {https://doi.org/10.1103/PhysRevLett.128.073604} {\bibfield  {journal} {\bibinfo  {journal} {Phys. Rev. Lett.}\ }\textbf {\bibinfo {volume} {128}},\ \bibinfo {pages} {073604} (\bibinfo {year} {2022})},\ \bibinfo {note} {and references therein}\BibitemShut {NoStop}%
\bibitem [{\citenamefont {Olivares}\ \emph {et~al.}(2003)\citenamefont {Olivares}, \citenamefont {Paris},\ and\ \citenamefont {Bonifacio}}]{Stefano:2003}%
  \BibitemOpen
  \bibfield  {author} {\bibinfo {author} {\bibfnamefont {S.}~\bibnamefont {Olivares}}, \bibinfo {author} {\bibfnamefont {M.~G.~A.}\ \bibnamefont {Paris}},\ and\ \bibinfo {author} {\bibfnamefont {R.}~\bibnamefont {Bonifacio}},\ }\bibfield  {title} {\bibinfo {title} {Teleportation improvement by inconclusive photon subtraction},\ }\href {https://doi.org/10.1103/PhysRevA.67.032314} {\bibfield  {journal} {\bibinfo  {journal} {Phys. Rev. A}\ }\textbf {\bibinfo {volume} {67}},\ \bibinfo {pages} {032314} (\bibinfo {year} {2003})}\BibitemShut {NoStop}%
\bibitem [{\citenamefont {Parigi}\ \emph {et~al.}(2007)\citenamefont {Parigi}, \citenamefont {Zavatta}, \citenamefont {Kim},\ and\ \citenamefont {Bellini}}]{Parigi:2007}%
  \BibitemOpen
  \bibfield  {author} {\bibinfo {author} {\bibfnamefont {V.}~\bibnamefont {Parigi}}, \bibinfo {author} {\bibfnamefont {A.}~\bibnamefont {Zavatta}}, \bibinfo {author} {\bibfnamefont {M.}~\bibnamefont {Kim}},\ and\ \bibinfo {author} {\bibfnamefont {M.}~\bibnamefont {Bellini}},\ }\bibfield  {title} {\bibinfo {title} {Probing quantum commutation rules by addition and subtraction of single photons to/from a light field},\ }\href {https://doi.org/10.1126/science.1146204} {\bibfield  {journal} {\bibinfo  {journal} {Science}\ }\textbf {\bibinfo {volume} {317}},\ \bibinfo {pages} {1890} (\bibinfo {year} {2007})}\BibitemShut {NoStop}%
\bibitem [{\citenamefont {Biswas}\ and\ \citenamefont {Agarwal}(2007)}]{Biswas:2007}%
  \BibitemOpen
  \bibfield  {author} {\bibinfo {author} {\bibfnamefont {A.}~\bibnamefont {Biswas}}\ and\ \bibinfo {author} {\bibfnamefont {G.~S.}\ \bibnamefont {Agarwal}},\ }\bibfield  {title} {\bibinfo {title} {Nonclassicality and decoherence of photon-subtracted squeezed states},\ }\href {https://doi.org/10.1103/PhysRevA.75.032104} {\bibfield  {journal} {\bibinfo  {journal} {Phys. Rev. A}\ }\textbf {\bibinfo {volume} {75}},\ \bibinfo {pages} {032104} (\bibinfo {year} {2007})}\BibitemShut {NoStop}%
\bibitem [{\citenamefont {Ourjoumtsev}\ \emph {et~al.}(2007)\citenamefont {Ourjoumtsev}, \citenamefont {Dantan}, \citenamefont {Tualle-Brouri},\ and\ \citenamefont {Grangier}}]{Grangier:2007}%
  \BibitemOpen
  \bibfield  {author} {\bibinfo {author} {\bibfnamefont {A.}~\bibnamefont {Ourjoumtsev}}, \bibinfo {author} {\bibfnamefont {A.}~\bibnamefont {Dantan}}, \bibinfo {author} {\bibfnamefont {R.}~\bibnamefont {Tualle-Brouri}},\ and\ \bibinfo {author} {\bibfnamefont {P.}~\bibnamefont {Grangier}},\ }\bibfield  {title} {\bibinfo {title} {Increasing entanglement between {G}aussian states by coherent photon subtraction},\ }\href {https://doi.org/10.1103/PhysRevLett.98.030502} {\bibfield  {journal} {\bibinfo  {journal} {Phys. Rev. Lett.}\ }\textbf {\bibinfo {volume} {98}},\ \bibinfo {pages} {030502} (\bibinfo {year} {2007})}\BibitemShut {NoStop}%
\bibitem [{\citenamefont {Takeoka}\ \emph {et~al.}(2008)\citenamefont {Takeoka}, \citenamefont {Takahashi},\ and\ \citenamefont {Sasaki}}]{Masahiro:2008}%
  \BibitemOpen
  \bibfield  {author} {\bibinfo {author} {\bibfnamefont {M.}~\bibnamefont {Takeoka}}, \bibinfo {author} {\bibfnamefont {H.}~\bibnamefont {Takahashi}},\ and\ \bibinfo {author} {\bibfnamefont {M.}~\bibnamefont {Sasaki}},\ }\bibfield  {title} {\bibinfo {title} {Large-amplitude coherent-state superposition generated by a time-separated two-photon subtraction from a continuous-wave squeezed vacuum},\ }\href {https://doi.org/10.1103/PhysRevA.77.062315} {\bibfield  {journal} {\bibinfo  {journal} {Phys. Rev. A}\ }\textbf {\bibinfo {volume} {77}},\ \bibinfo {pages} {062315} (\bibinfo {year} {2008})}\BibitemShut {NoStop}%
\bibitem [{\citenamefont {Xu}\ \emph {et~al.}(2012)\citenamefont {Xu}, \citenamefont {Yuan}, \citenamefont {Hu},\ and\ \citenamefont {Fan}}]{Xu:2012}%
  \BibitemOpen
  \bibfield  {author} {\bibinfo {author} {\bibfnamefont {X.-X.}\ \bibnamefont {Xu}}, \bibinfo {author} {\bibfnamefont {H.-C.}\ \bibnamefont {Yuan}}, \bibinfo {author} {\bibfnamefont {L.-Y.}\ \bibnamefont {Hu}},\ and\ \bibinfo {author} {\bibfnamefont {H.-Y.}\ \bibnamefont {Fan}},\ }\bibfield  {title} {\bibinfo {title} {Non-{G}aussianity of photon-added-then-subtracted squeezed vacuum state},\ }\href {https://doi.org/https://doi.org/10.1016/j.ijleo.2010.10.050} {\bibfield  {journal} {\bibinfo  {journal} {Optik}\ }\textbf {\bibinfo {volume} {123}},\ \bibinfo {pages} {16} (\bibinfo {year} {2012})}\BibitemShut {NoStop}%
\bibitem [{\citenamefont {Navarrete-Benlloch}\ \emph {et~al.}(2012)\citenamefont {Navarrete-Benlloch}, \citenamefont {Garc\'{\i}a-Patr\'on}, \citenamefont {Shapiro},\ and\ \citenamefont {Cerf}}]{Cerf:2012}%
  \BibitemOpen
  \bibfield  {author} {\bibinfo {author} {\bibfnamefont {C.}~\bibnamefont {Navarrete-Benlloch}}, \bibinfo {author} {\bibfnamefont {R.}~\bibnamefont {Garc\'{\i}a-Patr\'on}}, \bibinfo {author} {\bibfnamefont {J.~H.}\ \bibnamefont {Shapiro}},\ and\ \bibinfo {author} {\bibfnamefont {N.~J.}\ \bibnamefont {Cerf}},\ }\bibfield  {title} {\bibinfo {title} {Enhancing quantum entanglement by photon addition and subtraction},\ }\href {https://doi.org/10.1103/PhysRevA.86.012328} {\bibfield  {journal} {\bibinfo  {journal} {Phys. Rev. A}\ }\textbf {\bibinfo {volume} {86}},\ \bibinfo {pages} {012328} (\bibinfo {year} {2012})}\BibitemShut {NoStop}%
\bibitem [{\citenamefont {Das}\ \emph {et~al.}(2016)\citenamefont {Das}, \citenamefont {Prabhu}, \citenamefont {Sen(De)},\ and\ \citenamefont {Sen}}]{Das:2016}%
  \BibitemOpen
  \bibfield  {author} {\bibinfo {author} {\bibfnamefont {T.}~\bibnamefont {Das}}, \bibinfo {author} {\bibfnamefont {R.}~\bibnamefont {Prabhu}}, \bibinfo {author} {\bibfnamefont {A.}~\bibnamefont {Sen(De)}},\ and\ \bibinfo {author} {\bibfnamefont {U.}~\bibnamefont {Sen}},\ }\bibfield  {title} {\bibinfo {title} {Superiority of photon subtraction to addition for entanglement in a multimode squeezed vacuum},\ }\href {https://doi.org/10.1103/PhysRevA.93.052313} {\bibfield  {journal} {\bibinfo  {journal} {Phys. Rev. A}\ }\textbf {\bibinfo {volume} {93}},\ \bibinfo {pages} {052313} (\bibinfo {year} {2016})}\BibitemShut {NoStop}%
\bibitem [{\citenamefont {Barnett}\ \emph {et~al.}(2018)\citenamefont {Barnett}, \citenamefont {Ferenczi}, \citenamefont {Gilson},\ and\ \citenamefont {Speirits}}]{Barnett:2018}%
  \BibitemOpen
  \bibfield  {author} {\bibinfo {author} {\bibfnamefont {S.~M.}\ \bibnamefont {Barnett}}, \bibinfo {author} {\bibfnamefont {G.}~\bibnamefont {Ferenczi}}, \bibinfo {author} {\bibfnamefont {C.~R.}\ \bibnamefont {Gilson}},\ and\ \bibinfo {author} {\bibfnamefont {F.~C.}\ \bibnamefont {Speirits}},\ }\bibfield  {title} {\bibinfo {title} {Statistics of photon-subtracted and photon-added states},\ }\href {https://doi.org/10.1103/PhysRevA.98.013809} {\bibfield  {journal} {\bibinfo  {journal} {Phys. Rev. A}\ }\textbf {\bibinfo {volume} {98}},\ \bibinfo {pages} {013809} (\bibinfo {year} {2018})}\BibitemShut {NoStop}%
\bibitem [{\citenamefont {Takase}\ \emph {et~al.}(2021)\citenamefont {Takase}, \citenamefont {Yoshikawa}, \citenamefont {Asavanant}, \citenamefont {Endo},\ and\ \citenamefont {Furusawa}}]{Takase:2021}%
  \BibitemOpen
  \bibfield  {author} {\bibinfo {author} {\bibfnamefont {K.}~\bibnamefont {Takase}}, \bibinfo {author} {\bibfnamefont {J.-i.}\ \bibnamefont {Yoshikawa}}, \bibinfo {author} {\bibfnamefont {W.}~\bibnamefont {Asavanant}}, \bibinfo {author} {\bibfnamefont {M.}~\bibnamefont {Endo}},\ and\ \bibinfo {author} {\bibfnamefont {A.}~\bibnamefont {Furusawa}},\ }\bibfield  {title} {\bibinfo {title} {Generation of optical {S}chr\"odinger cat states by generalized photon subtraction},\ }\href {https://doi.org/10.1103/PhysRevA.103.013710} {\bibfield  {journal} {\bibinfo  {journal} {Phys. Rev. A}\ }\textbf {\bibinfo {volume} {103}},\ \bibinfo {pages} {013710} (\bibinfo {year} {2021})}\BibitemShut {NoStop}%
\bibitem [{\citenamefont {Grebien}\ \emph {et~al.}(2022)\citenamefont {Grebien}, \citenamefont {G\"ottsch}, \citenamefont {Hage}, \citenamefont {Fiur\'a\ifmmode~\check{s}\else \v{s}\fi{}ek},\ and\ \citenamefont {Schnabel}}]{Grebien:2022}%
  \BibitemOpen
  \bibfield  {author} {\bibinfo {author} {\bibfnamefont {S.}~\bibnamefont {Grebien}}, \bibinfo {author} {\bibfnamefont {J.}~\bibnamefont {G\"ottsch}}, \bibinfo {author} {\bibfnamefont {B.}~\bibnamefont {Hage}}, \bibinfo {author} {\bibfnamefont {J.}~\bibnamefont {Fiur\'a\ifmmode~\check{s}\else \v{s}\fi{}ek}},\ and\ \bibinfo {author} {\bibfnamefont {R.}~\bibnamefont {Schnabel}},\ }\bibfield  {title} {\bibinfo {title} {Multistep two-copy distillation of squeezed states via two-photon subtraction},\ }\href {https://doi.org/10.1103/PhysRevLett.129.273604} {\bibfield  {journal} {\bibinfo  {journal} {Phys. Rev. Lett.}\ }\textbf {\bibinfo {volume} {129}},\ \bibinfo {pages} {273604} (\bibinfo {year} {2022})}\BibitemShut {NoStop}%
\bibitem [{\citenamefont {Song}\ \emph {et~al.}(2023)\citenamefont {Song}, \citenamefont {Zhang},\ and\ \citenamefont {Yonezawa}}]{Hongbin:2023}%
  \BibitemOpen
  \bibfield  {author} {\bibinfo {author} {\bibfnamefont {H.}~\bibnamefont {Song}}, \bibinfo {author} {\bibfnamefont {G.}~\bibnamefont {Zhang}},\ and\ \bibinfo {author} {\bibfnamefont {H.}~\bibnamefont {Yonezawa}},\ }\bibfield  {title} {\bibinfo {title} {Strong quantum entanglement based on two-mode photon-subtracted squeezed vacuum states},\ }\href {https://doi.org/10.1103/PhysRevA.108.052420} {\bibfield  {journal} {\bibinfo  {journal} {Phys. Rev. A}\ }\textbf {\bibinfo {volume} {108}},\ \bibinfo {pages} {052420} (\bibinfo {year} {2023})}\BibitemShut {NoStop}%
\bibitem [{\citenamefont {Tomoda}\ \emph {et~al.}(2024)\citenamefont {Tomoda}, \citenamefont {Machinaga}, \citenamefont {Takase}, \citenamefont {Harada}, \citenamefont {Kashiwazaki}, \citenamefont {Umeki}, \citenamefont {Miki}, \citenamefont {China}, \citenamefont {Yabuno}, \citenamefont {Terai}, \citenamefont {Okuno},\ and\ \citenamefont {Takeda}}]{Tomoda:2024}%
  \BibitemOpen
  \bibfield  {author} {\bibinfo {author} {\bibfnamefont {H.}~\bibnamefont {Tomoda}}, \bibinfo {author} {\bibfnamefont {A.}~\bibnamefont {Machinaga}}, \bibinfo {author} {\bibfnamefont {K.}~\bibnamefont {Takase}}, \bibinfo {author} {\bibfnamefont {J.}~\bibnamefont {Harada}}, \bibinfo {author} {\bibfnamefont {T.}~\bibnamefont {Kashiwazaki}}, \bibinfo {author} {\bibfnamefont {T.}~\bibnamefont {Umeki}}, \bibinfo {author} {\bibfnamefont {S.}~\bibnamefont {Miki}}, \bibinfo {author} {\bibfnamefont {F.}~\bibnamefont {China}}, \bibinfo {author} {\bibfnamefont {M.}~\bibnamefont {Yabuno}}, \bibinfo {author} {\bibfnamefont {H.}~\bibnamefont {Terai}}, \bibinfo {author} {\bibfnamefont {D.}~\bibnamefont {Okuno}},\ and\ \bibinfo {author} {\bibfnamefont {S.}~\bibnamefont {Takeda}},\ }\bibfield  {title} {\bibinfo {title} {Boosting generation rate of squeezed single-photon states by generalized photon subtraction},\ }\bibfield  {journal} {\bibinfo  {journal} {arXiv:2404.19304 [quant-ph]}\ }\href
  {https://doi.org/10.48550/arXiv.2404.19304} {10.48550/arXiv.2404.19304} (\bibinfo {year} {2024})\BibitemShut {NoStop}%
\bibitem [{\citenamefont {Fadrn{\'y}}\ \emph {et~al.}(2024)\citenamefont {Fadrn{\'y}}, \citenamefont {Neset}, \citenamefont {Bielak}, \citenamefont {Je{\v{z}}ek}, \citenamefont {B{\'i}lek},\ and\ \citenamefont {Fiur{\'a}{\v{s}}ek}}]{Fadrny:2024}%
  \BibitemOpen
  \bibfield  {author} {\bibinfo {author} {\bibfnamefont {J.}~\bibnamefont {Fadrn{\'y}}}, \bibinfo {author} {\bibfnamefont {M.}~\bibnamefont {Neset}}, \bibinfo {author} {\bibfnamefont {M.}~\bibnamefont {Bielak}}, \bibinfo {author} {\bibfnamefont {M.}~\bibnamefont {Je{\v{z}}ek}}, \bibinfo {author} {\bibfnamefont {J.}~\bibnamefont {B{\'i}lek}},\ and\ \bibinfo {author} {\bibfnamefont {J.}~\bibnamefont {Fiur{\'a}{\v{s}}ek}},\ }\bibfield  {title} {\bibinfo {title} {Experimental preparation of multiphoton-added coherent states of light},\ }\href {https://doi.org/10.1038/s41534-024-00885-y} {\bibfield  {journal} {\bibinfo  {journal} {npj Quantum Information}\ }\textbf {\bibinfo {volume} {10}},\ \bibinfo {pages} {89} (\bibinfo {year} {2024})}\BibitemShut {NoStop}%
\bibitem [{\citenamefont {Kim}(2008)}]{Kim:2008}%
  \BibitemOpen
  \bibfield  {author} {\bibinfo {author} {\bibfnamefont {M.~S.}\ \bibnamefont {Kim}},\ }\bibfield  {title} {\bibinfo {title} {Recent developments in photon-level operations on travelling light fields},\ }\href {https://doi.org/10.1088/0953-4075/41/13/133001} {\bibfield  {journal} {\bibinfo  {journal} {Journal of Physics B: Atomic, Molecular and Optical Physics}\ }\textbf {\bibinfo {volume} {41}},\ \bibinfo {pages} {133001} (\bibinfo {year} {2008})}\BibitemShut {NoStop}%
\bibitem [{\citenamefont {Fiur\'a\ifmmode~\check{s}\else \v{s}\fi{}ek}(2009)}]{Fiurasek:2009}%
  \BibitemOpen
  \bibfield  {author} {\bibinfo {author} {\bibfnamefont {J.}~\bibnamefont {Fiur\'a\ifmmode~\check{s}\else \v{s}\fi{}ek}},\ }\bibfield  {title} {\bibinfo {title} {Engineering quantum operations on traveling light beams by multiple photon addition and subtraction},\ }\href {https://doi.org/10.1103/PhysRevA.80.053822} {\bibfield  {journal} {\bibinfo  {journal} {Phys. Rev. A}\ }\textbf {\bibinfo {volume} {80}},\ \bibinfo {pages} {053822} (\bibinfo {year} {2009})}\BibitemShut {NoStop}%
\bibitem [{\citenamefont {Zhang}\ and\ \citenamefont {Fan}(1992)}]{Zhang:1992}%
  \BibitemOpen
  \bibfield  {author} {\bibinfo {author} {\bibfnamefont {Z.}~\bibnamefont {Zhang}}\ and\ \bibinfo {author} {\bibfnamefont {H.}~\bibnamefont {Fan}},\ }\bibfield  {title} {\bibinfo {title} {Properties of states generated by excitations on a squeezed vacuum state},\ }\href {https://doi.org/https://doi.org/10.1016/0375-9601(92)91046-T} {\bibfield  {journal} {\bibinfo  {journal} {Physics Letters A}\ }\textbf {\bibinfo {volume} {165}},\ \bibinfo {pages} {14} (\bibinfo {year} {1992})}\BibitemShut {NoStop}%
\bibitem [{\citenamefont {Agarwal}\ and\ \citenamefont {Tara}(1991)}]{Agarwal:1991}%
  \BibitemOpen
  \bibfield  {author} {\bibinfo {author} {\bibfnamefont {G.~S.}\ \bibnamefont {Agarwal}}\ and\ \bibinfo {author} {\bibfnamefont {K.}~\bibnamefont {Tara}},\ }\bibfield  {title} {\bibinfo {title} {Nonclassical properties of states generated by the excitations on a coherent state},\ }\href {https://doi.org/10.1103/PhysRevA.43.492} {\bibfield  {journal} {\bibinfo  {journal} {Phys. Rev. A}\ }\textbf {\bibinfo {volume} {43}},\ \bibinfo {pages} {492} (\bibinfo {year} {1991})}\BibitemShut {NoStop}%
\bibitem [{\citenamefont {Thapliyal}\ \emph {et~al.}(2024)\citenamefont {Thapliyal}, \citenamefont {Jr.}, \citenamefont {Haderka}, \citenamefont {Mich\'{a}lek},\ and\ \citenamefont {Machulka}}]{Thapliyal:2024}%
  \BibitemOpen
  \bibfield  {author} {\bibinfo {author} {\bibfnamefont {K.}~\bibnamefont {Thapliyal}}, \bibinfo {author} {\bibfnamefont {J.~P.}\ \bibnamefont {Jr.}}, \bibinfo {author} {\bibfnamefont {O.}~\bibnamefont {Haderka}}, \bibinfo {author} {\bibfnamefont {V.}~\bibnamefont {Mich\'{a}lek}},\ and\ \bibinfo {author} {\bibfnamefont {R.}~\bibnamefont {Machulka}},\ }\bibfield  {title} {\bibinfo {title} {Experimental photon addition and subtraction in multi-mode and entangled optical fields},\ }\href {https://doi.org/10.1364/OL.532242} {\bibfield  {journal} {\bibinfo  {journal} {Opt. Lett.}\ }\textbf {\bibinfo {volume} {49}},\ \bibinfo {pages} {4521} (\bibinfo {year} {2024})}\BibitemShut {NoStop}%
\bibitem [{\citenamefont {Pe\v{r}ina}\ \emph {et~al.}(2024)\citenamefont {Pe\v{r}ina}, \citenamefont {Thapliyal}, \citenamefont {Haderka}, \citenamefont {Mich\'{a}lek},\ and\ \citenamefont {Machulka}}]{Perina:2024}%
  \BibitemOpen
  \bibfield  {author} {\bibinfo {author} {\bibfnamefont {J.}~\bibnamefont {Pe\v{r}ina}}, \bibinfo {author} {\bibfnamefont {K.}~\bibnamefont {Thapliyal}}, \bibinfo {author} {\bibfnamefont {O.}~\bibnamefont {Haderka}}, \bibinfo {author} {\bibfnamefont {V.}~\bibnamefont {Mich\'{a}lek}},\ and\ \bibinfo {author} {\bibfnamefont {R.}~\bibnamefont {Machulka}},\ }\bibfield  {title} {\bibinfo {title} {Sub-{P}oissonian twin beams},\ }\href {https://doi.org/10.1364/OPTICAQ.509228} {\bibfield  {journal} {\bibinfo  {journal} {Optica Quantum}\ }\textbf {\bibinfo {volume} {2}},\ \bibinfo {pages} {148} (\bibinfo {year} {2024})}\BibitemShut {NoStop}%
\bibitem [{\citenamefont {Walschaers}(2021)}]{Walschaers:2021}%
  \BibitemOpen
  \bibfield  {author} {\bibinfo {author} {\bibfnamefont {M.}~\bibnamefont {Walschaers}},\ }\bibfield  {title} {\bibinfo {title} {Non-{G}aussian quantum states and where to find them},\ }\href {https://doi.org/10.1103/PRXQuantum.2.030204} {\bibfield  {journal} {\bibinfo  {journal} {PRX Quantum}\ }\textbf {\bibinfo {volume} {2}},\ \bibinfo {pages} {030204} (\bibinfo {year} {2021})}\BibitemShut {NoStop}%
\bibitem [{\citenamefont {Biagi}\ \emph {et~al.}(2022)\citenamefont {Biagi}, \citenamefont {Francesconi}, \citenamefont {Zavatta},\ and\ \citenamefont {Bellini}}]{Biagi:2022}%
  \BibitemOpen
  \bibfield  {author} {\bibinfo {author} {\bibfnamefont {N.}~\bibnamefont {Biagi}}, \bibinfo {author} {\bibfnamefont {S.}~\bibnamefont {Francesconi}}, \bibinfo {author} {\bibfnamefont {A.}~\bibnamefont {Zavatta}},\ and\ \bibinfo {author} {\bibfnamefont {M.}~\bibnamefont {Bellini}},\ }\bibfield  {title} {\bibinfo {title} {Photon-by-photon quantum light state engineering},\ }\href {https://doi.org/https://doi.org/10.1016/j.pquantelec.2022.100414} {\bibfield  {journal} {\bibinfo  {journal} {Progress in Quantum Electronics}\ }\textbf {\bibinfo {volume} {84}},\ \bibinfo {pages} {100414} (\bibinfo {year} {2022})}\BibitemShut {NoStop}%
\bibitem [{\citenamefont {G\'orecki}\ \emph {et~al.}(2022)\citenamefont {G\'orecki}, \citenamefont {Riccardi},\ and\ \citenamefont {Maccone}}]{Gorecki:2022}%
  \BibitemOpen
  \bibfield  {author} {\bibinfo {author} {\bibfnamefont {W.}~\bibnamefont {G\'orecki}}, \bibinfo {author} {\bibfnamefont {A.}~\bibnamefont {Riccardi}},\ and\ \bibinfo {author} {\bibfnamefont {L.}~\bibnamefont {Maccone}},\ }\bibfield  {title} {\bibinfo {title} {Quantum metrology of noisy spreading channels},\ }\href {https://doi.org/10.1103/PhysRevLett.129.240503} {\bibfield  {journal} {\bibinfo  {journal} {Phys. Rev. Lett.}\ }\textbf {\bibinfo {volume} {129}},\ \bibinfo {pages} {240503} (\bibinfo {year} {2022})}\BibitemShut {NoStop}%
\bibitem [{\citenamefont {Lloyd}\ and\ \citenamefont {Braunstein}(1999)}]{Lloyd:1999}%
  \BibitemOpen
  \bibfield  {author} {\bibinfo {author} {\bibfnamefont {S.}~\bibnamefont {Lloyd}}\ and\ \bibinfo {author} {\bibfnamefont {S.~L.}\ \bibnamefont {Braunstein}},\ }\bibfield  {title} {\bibinfo {title} {Quantum computation over continuous variables},\ }\href {https://doi.org/10.1103/PhysRevLett.82.1784} {\bibfield  {journal} {\bibinfo  {journal} {Phys. Rev. Lett.}\ }\textbf {\bibinfo {volume} {82}},\ \bibinfo {pages} {1784} (\bibinfo {year} {1999})}\BibitemShut {NoStop}%
\bibitem [{\citenamefont {Menicucci}\ \emph {et~al.}(2006)\citenamefont {Menicucci}, \citenamefont {Van~Loock}, \citenamefont {Gu}, \citenamefont {Weedbrook}, \citenamefont {Ralph},\ and\ \citenamefont {Nielsen}}]{Menicucci:2006}%
  \BibitemOpen
  \bibfield  {author} {\bibinfo {author} {\bibfnamefont {N.~C.}\ \bibnamefont {Menicucci}}, \bibinfo {author} {\bibfnamefont {P.}~\bibnamefont {Van~Loock}}, \bibinfo {author} {\bibfnamefont {M.}~\bibnamefont {Gu}}, \bibinfo {author} {\bibfnamefont {C.}~\bibnamefont {Weedbrook}}, \bibinfo {author} {\bibfnamefont {T.~C.}\ \bibnamefont {Ralph}},\ and\ \bibinfo {author} {\bibfnamefont {M.~A.}\ \bibnamefont {Nielsen}},\ }\bibfield  {title} {\bibinfo {title} {Universal quantum computation with continuous-variable cluster states},\ }\href {https://doi.org/10.1103/PhysRevLett.97.110501} {\bibfield  {journal} {\bibinfo  {journal} {Phys. Rev. Lett.}\ }\textbf {\bibinfo {volume} {97}},\ \bibinfo {pages} {110501} (\bibinfo {year} {2006})}\BibitemShut {NoStop}%
\bibitem [{\citenamefont {Bartlett}\ \emph {et~al.}(2002)\citenamefont {Bartlett}, \citenamefont {Sanders}, \citenamefont {Braunstein},\ and\ \citenamefont {Nemoto}}]{Bartlett:2002}%
  \BibitemOpen
  \bibfield  {author} {\bibinfo {author} {\bibfnamefont {S.~D.}\ \bibnamefont {Bartlett}}, \bibinfo {author} {\bibfnamefont {B.~C.}\ \bibnamefont {Sanders}}, \bibinfo {author} {\bibfnamefont {S.~L.}\ \bibnamefont {Braunstein}},\ and\ \bibinfo {author} {\bibfnamefont {K.}~\bibnamefont {Nemoto}},\ }\bibfield  {title} {\bibinfo {title} {Efficient classical simulation of continuous variable quantum information processes},\ }\href {https://doi.org/10.1103/PhysRevLett.88.097904} {\bibfield  {journal} {\bibinfo  {journal} {Phys. Rev. Lett.}\ }\textbf {\bibinfo {volume} {88}},\ \bibinfo {pages} {097904} (\bibinfo {year} {2002})}\BibitemShut {NoStop}%
\bibitem [{\citenamefont {Mari}\ and\ \citenamefont {Eisert}(2012)}]{Mari:2012}%
  \BibitemOpen
  \bibfield  {author} {\bibinfo {author} {\bibfnamefont {A.}~\bibnamefont {Mari}}\ and\ \bibinfo {author} {\bibfnamefont {J.}~\bibnamefont {Eisert}},\ }\bibfield  {title} {\bibinfo {title} {Positive wigner functions render classical simulation of quantum computation efficient},\ }\href {https://doi.org/10.1103/PhysRevLett.109.230503} {\bibfield  {journal} {\bibinfo  {journal} {Phys. Rev. Lett.}\ }\textbf {\bibinfo {volume} {109}},\ \bibinfo {pages} {230503} (\bibinfo {year} {2012})}\BibitemShut {NoStop}%
\bibitem [{\citenamefont {Olson}\ \emph {et~al.}(2015)\citenamefont {Olson}, \citenamefont {Seshadreesan}, \citenamefont {Motes}, \citenamefont {Rohde},\ and\ \citenamefont {Dowling}}]{Olson:2015}%
  \BibitemOpen
  \bibfield  {author} {\bibinfo {author} {\bibfnamefont {J.~P.}\ \bibnamefont {Olson}}, \bibinfo {author} {\bibfnamefont {K.~P.}\ \bibnamefont {Seshadreesan}}, \bibinfo {author} {\bibfnamefont {K.~R.}\ \bibnamefont {Motes}}, \bibinfo {author} {\bibfnamefont {P.~P.}\ \bibnamefont {Rohde}},\ and\ \bibinfo {author} {\bibfnamefont {J.~P.}\ \bibnamefont {Dowling}},\ }\bibfield  {title} {\bibinfo {title} {Sampling arbitrary photon-added or photon-subtracted squeezed states is in the same complexity class as boson sampling},\ }\href {https://doi.org/10.1103/PhysRevA.91.022317} {\bibfield  {journal} {\bibinfo  {journal} {Phys. Rev. A}\ }\textbf {\bibinfo {volume} {91}},\ \bibinfo {pages} {022317} (\bibinfo {year} {2015})}\BibitemShut {NoStop}%
\bibitem [{\citenamefont {Quesne}(2001)}]{Quesne:2001}%
  \BibitemOpen
  \bibfield  {author} {\bibinfo {author} {\bibfnamefont {C.}~\bibnamefont {Quesne}},\ }\bibfield  {title} {\bibinfo {title} {Completeness of photon-added squeezed vacuum and one-photon states and of photon-added coherent states on a circle},\ }\href {https://doi.org/https://doi.org/10.1016/S0375-9601(01)00554-0} {\bibfield  {journal} {\bibinfo  {journal} {Physics Letters A}\ }\textbf {\bibinfo {volume} {288}},\ \bibinfo {pages} {241} (\bibinfo {year} {2001})}\BibitemShut {NoStop}%
\bibitem [{\citenamefont {Yuan}\ \emph {et~al.}(2019)\citenamefont {Yuan}, \citenamefont {Xu}, \citenamefont {Cai},\ and\ \citenamefont {Xu}}]{Yuan:2019}%
  \BibitemOpen
  \bibfield  {author} {\bibinfo {author} {\bibfnamefont {H.-C.}\ \bibnamefont {Yuan}}, \bibinfo {author} {\bibfnamefont {X.-X.}\ \bibnamefont {Xu}}, \bibinfo {author} {\bibfnamefont {J.-W.}\ \bibnamefont {Cai}},\ and\ \bibinfo {author} {\bibfnamefont {Y.-J.}\ \bibnamefont {Xu}},\ }\bibfield  {title} {\bibinfo {title} {Single-mode squeezed vacuum state orthogonalization via photon-addition operation},\ }\href {https://doi.org/https://doi.org/10.1016/j.ijleo.2019.02.120} {\bibfield  {journal} {\bibinfo  {journal} {Optik}\ }\textbf {\bibinfo {volume} {183}},\ \bibinfo {pages} {1043} (\bibinfo {year} {2019})}\BibitemShut {NoStop}%
\bibitem [{\citenamefont {Bohloul}\ \emph {et~al.}(2024)\citenamefont {Bohloul}, \citenamefont {Dehghani},\ and\ \citenamefont {Fakhri}}]{Bohloul:2024}%
  \BibitemOpen
  \bibfield  {author} {\bibinfo {author} {\bibfnamefont {M.}~\bibnamefont {Bohloul}}, \bibinfo {author} {\bibfnamefont {A.}~\bibnamefont {Dehghani}},\ and\ \bibinfo {author} {\bibfnamefont {H.}~\bibnamefont {Fakhri}},\ }\bibfield  {title} {\bibinfo {title} {Generating superposition of squeezed states and photon-added squeezed states},\ }\href {https://doi.org/10.1088/1402-4896/ad2f8e} {\bibfield  {journal} {\bibinfo  {journal} {Physica Scripta}\ }\textbf {\bibinfo {volume} {99}},\ \bibinfo {pages} {045112} (\bibinfo {year} {2024})}\BibitemShut {NoStop}%
\bibitem [{\citenamefont {Guo}\ \emph {et~al.}(2018)\citenamefont {Guo}, \citenamefont {Yu},\ and\ \citenamefont {Zhang}}]{Guo:2018}%
  \BibitemOpen
  \bibfield  {author} {\bibinfo {author} {\bibfnamefont {L.-L.}\ \bibnamefont {Guo}}, \bibinfo {author} {\bibfnamefont {Y.-F.}\ \bibnamefont {Yu}},\ and\ \bibinfo {author} {\bibfnamefont {Z.-M.}\ \bibnamefont {Zhang}},\ }\bibfield  {title} {\bibinfo {title} {Improving the phase sensitivity of an {SU}(1,1) interferometer with photon-added squeezed vacuum light},\ }\href {https://doi.org/10.1364/OE.26.029099} {\bibfield  {journal} {\bibinfo  {journal} {Opt. Express}\ }\textbf {\bibinfo {volume} {26}},\ \bibinfo {pages} {29099} (\bibinfo {year} {2018})}\BibitemShut {NoStop}%
\bibitem [{\citenamefont {Chen}\ \emph {et~al.}(2024)\citenamefont {Chen}, \citenamefont {Hsieh}, \citenamefont {Ning}, \citenamefont {Wu}, \citenamefont {Chen}, \citenamefont {Shi}, \citenamefont {Yang}, \citenamefont {Steuernagel}, \citenamefont {Wu},\ and\ \citenamefont {Lee}}]{Chen:2024}%
  \BibitemOpen
  \bibfield  {author} {\bibinfo {author} {\bibfnamefont {Y.-R.}\ \bibnamefont {Chen}}, \bibinfo {author} {\bibfnamefont {H.-Y.}\ \bibnamefont {Hsieh}}, \bibinfo {author} {\bibfnamefont {J.}~\bibnamefont {Ning}}, \bibinfo {author} {\bibfnamefont {H.-C.}\ \bibnamefont {Wu}}, \bibinfo {author} {\bibfnamefont {H.~L.}\ \bibnamefont {Chen}}, \bibinfo {author} {\bibfnamefont {Z.-H.}\ \bibnamefont {Shi}}, \bibinfo {author} {\bibfnamefont {P.}~\bibnamefont {Yang}}, \bibinfo {author} {\bibfnamefont {O.}~\bibnamefont {Steuernagel}}, \bibinfo {author} {\bibfnamefont {C.-M.}\ \bibnamefont {Wu}},\ and\ \bibinfo {author} {\bibfnamefont {R.-K.}\ \bibnamefont {Lee}},\ }\bibfield  {title} {\bibinfo {title} {Generation of heralded optical cat states by photon addition},\ }\href {https://doi.org/10.1103/PhysRevA.110.023703} {\bibfield  {journal} {\bibinfo  {journal} {Phys. Rev. A}\ }\textbf {\bibinfo {volume} {110}},\ \bibinfo {pages} {023703} (\bibinfo {year} {2024})}\BibitemShut {NoStop}%
\bibitem [{\citenamefont {Neergaard-Nielsen}\ \emph {et~al.}(2006)\citenamefont {Neergaard-Nielsen}, \citenamefont {Nielsen}, \citenamefont {Hettich}, \citenamefont {M\o{}lmer},\ and\ \citenamefont {Polzik}}]{Neergaard-Nielsen:2006}%
  \BibitemOpen
  \bibfield  {author} {\bibinfo {author} {\bibfnamefont {J.~S.}\ \bibnamefont {Neergaard-Nielsen}}, \bibinfo {author} {\bibfnamefont {B.~M.}\ \bibnamefont {Nielsen}}, \bibinfo {author} {\bibfnamefont {C.}~\bibnamefont {Hettich}}, \bibinfo {author} {\bibfnamefont {K.}~\bibnamefont {M\o{}lmer}},\ and\ \bibinfo {author} {\bibfnamefont {E.~S.}\ \bibnamefont {Polzik}},\ }\bibfield  {title} {\bibinfo {title} {Generation of a superposition of odd photon number states for quantum information networks},\ }\href {https://doi.org/10.1103/PhysRevLett.97.083604} {\bibfield  {journal} {\bibinfo  {journal} {Phys. Rev. Lett.}\ }\textbf {\bibinfo {volume} {97}},\ \bibinfo {pages} {083604} (\bibinfo {year} {2006})}\BibitemShut {NoStop}%
\bibitem [{\citenamefont {Wakui}\ \emph {et~al.}(2007)\citenamefont {Wakui}, \citenamefont {Takahashi}, \citenamefont {Furusawa},\ and\ \citenamefont {Sasaki}}]{Wakui:2007}%
  \BibitemOpen
  \bibfield  {author} {\bibinfo {author} {\bibfnamefont {K.}~\bibnamefont {Wakui}}, \bibinfo {author} {\bibfnamefont {H.}~\bibnamefont {Takahashi}}, \bibinfo {author} {\bibfnamefont {A.}~\bibnamefont {Furusawa}},\ and\ \bibinfo {author} {\bibfnamefont {M.}~\bibnamefont {Sasaki}},\ }\bibfield  {title} {\bibinfo {title} {Photon subtracted squeezed states generated with periodically poled {KTiOPO{$_4$}}},\ }\href {https://doi.org/10.1364/OE.15.003568} {\bibfield  {journal} {\bibinfo  {journal} {Opt. Express}\ }\textbf {\bibinfo {volume} {15}},\ \bibinfo {pages} {3568} (\bibinfo {year} {2007})}\BibitemShut {NoStop}%
\bibitem [{\citenamefont {Namekata}\ \emph {et~al.}(2010)\citenamefont {Namekata}, \citenamefont {Takahashi}, \citenamefont {Fujii}, \citenamefont {Fukuda}, \citenamefont {Kurimura},\ and\ \citenamefont {Inoue}}]{Namekata:2010}%
  \BibitemOpen
  \bibfield  {author} {\bibinfo {author} {\bibfnamefont {N.}~\bibnamefont {Namekata}}, \bibinfo {author} {\bibfnamefont {Y.}~\bibnamefont {Takahashi}}, \bibinfo {author} {\bibfnamefont {G.}~\bibnamefont {Fujii}}, \bibinfo {author} {\bibfnamefont {D.}~\bibnamefont {Fukuda}}, \bibinfo {author} {\bibfnamefont {S.}~\bibnamefont {Kurimura}},\ and\ \bibinfo {author} {\bibfnamefont {S.}~\bibnamefont {Inoue}},\ }\bibfield  {title} {\bibinfo {title} {Non-{G}aussian operation based on photon subtraction using a photon-number-resolving detector at a telecommunications wavelength},\ }\href {https://doi.org/10.1038/nphoton.2010.158} {\bibfield  {journal} {\bibinfo  {journal} {Nature Photonics}\ }\textbf {\bibinfo {volume} {4}},\ \bibinfo {pages} {655} (\bibinfo {year} {2010})}\BibitemShut {NoStop}%
\bibitem [{\citenamefont {Gerrits}\ \emph {et~al.}(2010)\citenamefont {Gerrits}, \citenamefont {Glancy}, \citenamefont {Clement}, \citenamefont {Calkins}, \citenamefont {Lita}, \citenamefont {Miller}, \citenamefont {Migdall}, \citenamefont {Nam}, \citenamefont {Mirin},\ and\ \citenamefont {Knill}}]{Gerrits:2010}%
  \BibitemOpen
  \bibfield  {author} {\bibinfo {author} {\bibfnamefont {T.}~\bibnamefont {Gerrits}}, \bibinfo {author} {\bibfnamefont {S.}~\bibnamefont {Glancy}}, \bibinfo {author} {\bibfnamefont {T.~S.}\ \bibnamefont {Clement}}, \bibinfo {author} {\bibfnamefont {B.}~\bibnamefont {Calkins}}, \bibinfo {author} {\bibfnamefont {A.~E.}\ \bibnamefont {Lita}}, \bibinfo {author} {\bibfnamefont {A.~J.}\ \bibnamefont {Miller}}, \bibinfo {author} {\bibfnamefont {A.~L.}\ \bibnamefont {Migdall}}, \bibinfo {author} {\bibfnamefont {S.~W.}\ \bibnamefont {Nam}}, \bibinfo {author} {\bibfnamefont {R.~P.}\ \bibnamefont {Mirin}},\ and\ \bibinfo {author} {\bibfnamefont {E.}~\bibnamefont {Knill}},\ }\bibfield  {title} {\bibinfo {title} {Generation of optical coherent-state superpositions by number-resolved photon subtraction from the squeezed vacuum},\ }\href {https://doi.org/10.1103/PhysRevA.82.031802} {\bibfield  {journal} {\bibinfo  {journal} {Phys. Rev. A}\ }\textbf {\bibinfo {volume} {82}},\ \bibinfo {pages} {031802} (\bibinfo {year}
  {2010})}\BibitemShut {NoStop}%
\bibitem [{\citenamefont {Vaserstein}(1969)}]{Vaserstein:1969}%
  \BibitemOpen
  \bibfield  {author} {\bibinfo {author} {\bibfnamefont {L.~N.}\ \bibnamefont {Vaserstein}},\ }\href@noop {} {\bibfield  {journal} {\bibinfo  {journal} {Probl. Peredachi Inf.}\ }\textbf {\bibinfo {volume} {5}},\ \bibinfo {pages} {64} (\bibinfo {year} {1969})}\BibitemShut {NoStop}%
\bibitem [{\citenamefont {Kullback}\ and\ \citenamefont {Leibler}(1951)}]{KL:1951}%
  \BibitemOpen
  \bibfield  {author} {\bibinfo {author} {\bibfnamefont {R.}~\bibnamefont {Kullback}}\ and\ \bibinfo {author} {\bibfnamefont {R.~A.}\ \bibnamefont {Leibler}},\ }\bibfield  {title} {\bibinfo {title} {On information and sufficiency},\ }\href {https://doi.org/10.1214/aoms/1177729694} {\bibfield  {journal} {\bibinfo  {journal} {Ann. Math. Statist.}\ }\textbf {\bibinfo {volume} {22}},\ \bibinfo {pages} {79} (\bibinfo {year} {1951})}\BibitemShut {NoStop}%
\bibitem [{\citenamefont {Bhattacharyya}(1943)}]{Bhattacharyya:1943}%
  \BibitemOpen
  \bibfield  {author} {\bibinfo {author} {\bibfnamefont {A.}~\bibnamefont {Bhattacharyya}},\ }\href@noop {} {\bibfield  {journal} {\bibinfo  {journal} {Bull. Cal. Math. Soc.}\ }\textbf {\bibinfo {volume} {35}},\ \bibinfo {pages} {99} (\bibinfo {year} {1943})}\BibitemShut {NoStop}%
\bibitem [{\citenamefont {Sharmila}\ \emph {et~al.}(2019)\citenamefont {Sharmila}, \citenamefont {Lakshmibala},\ and\ \citenamefont {Balakrishnan}}]{Sharmila:2019}%
  \BibitemOpen
  \bibfield  {author} {\bibinfo {author} {\bibfnamefont {B.}~\bibnamefont {Sharmila}}, \bibinfo {author} {\bibfnamefont {S.}~\bibnamefont {Lakshmibala}},\ and\ \bibinfo {author} {\bibfnamefont {V.}~\bibnamefont {Balakrishnan}},\ }\bibfield  {title} {\bibinfo {title} {Estimation of entanglement in bipartite systems directly from tomograms},\ }\href {https://doi.org/10.1007/s11128-019-2352-0} {\bibfield  {journal} {\bibinfo  {journal} {Quantum Information Processing}\ }\textbf {\bibinfo {volume} {18}},\ \bibinfo {pages} {236} (\bibinfo {year} {2019})}\BibitemShut {NoStop}%
\bibitem [{\citenamefont {Paul}\ \emph {et~al.}(2024)\citenamefont {Paul}, \citenamefont {Ramanan}, \citenamefont {Balakrishnan},\ and\ \citenamefont {Lakshmibala}}]{Soumyabrata:2024}%
  \BibitemOpen
  \bibfield  {author} {\bibinfo {author} {\bibfnamefont {S.}~\bibnamefont {Paul}}, \bibinfo {author} {\bibfnamefont {S.}~\bibnamefont {Ramanan}}, \bibinfo {author} {\bibfnamefont {V.}~\bibnamefont {Balakrishnan}},\ and\ \bibinfo {author} {\bibfnamefont {S.}~\bibnamefont {Lakshmibala}},\ }\bibfield  {title} {\bibinfo {title} {Wasserstein distance and entropic divergences between quantum states of light},\ }\bibfield  {journal} {\bibinfo  {journal} {arXiv:2401.16098 [quant-ph]}\ }\href {https://doi.org/10.48550/arXiv.2401.16098} {10.48550/arXiv.2401.16098} (\bibinfo {year} {2024})\BibitemShut {NoStop}%
\bibitem [{\citenamefont {Shanta}\ \emph {et~al.}(1994)\citenamefont {Shanta}, \citenamefont {Chaturvedi}, \citenamefont {Srinivasan}, \citenamefont {Agarwal},\ and\ \citenamefont {Mehta}}]{Shanta:1994}%
  \BibitemOpen
  \bibfield  {author} {\bibinfo {author} {\bibfnamefont {P.}~\bibnamefont {Shanta}}, \bibinfo {author} {\bibfnamefont {S.}~\bibnamefont {Chaturvedi}}, \bibinfo {author} {\bibfnamefont {V.}~\bibnamefont {Srinivasan}}, \bibinfo {author} {\bibfnamefont {G.~S.}\ \bibnamefont {Agarwal}},\ and\ \bibinfo {author} {\bibfnamefont {C.~L.}\ \bibnamefont {Mehta}},\ }\bibfield  {title} {\bibinfo {title} {Unified approach to multiphoton coherent states},\ }\href {https://doi.org/10.1103/PhysRevLett.72.1447} {\bibfield  {journal} {\bibinfo  {journal} {Phys. Rev. Lett.}\ }\textbf {\bibinfo {volume} {72}},\ \bibinfo {pages} {1447} (\bibinfo {year} {1994})}\BibitemShut {NoStop}%
\bibitem [{\citenamefont {Laha}\ \emph {et~al.}(2018)\citenamefont {Laha}, \citenamefont {Lakshmibala},\ and\ \citenamefont {Balakrishnan}}]{Laha:2018}%
  \BibitemOpen
  \bibfield  {author} {\bibinfo {author} {\bibfnamefont {P.}~\bibnamefont {Laha}}, \bibinfo {author} {\bibfnamefont {S.}~\bibnamefont {Lakshmibala}},\ and\ \bibinfo {author} {\bibfnamefont {V.}~\bibnamefont {Balakrishnan}},\ }\bibfield  {title} {\bibinfo {title} {Estimation of nonclassical properties of multiphoton coherent states from optical tomograms},\ }\href {https://doi.org/10.1080/09500340.2018.1454527} {\bibfield  {journal} {\bibinfo  {journal} {Journal of Modern Optics}\ }\textbf {\bibinfo {volume} {65}},\ \bibinfo {pages} {1466} (\bibinfo {year} {2018})}\BibitemShut {NoStop}%
\bibitem [{\citenamefont {Ibort}\ \emph {et~al.}(2009)\citenamefont {Ibort}, \citenamefont {Man{\textquotesingle}ko}, \citenamefont {Marmo}, \citenamefont {Simoni},\ and\ \citenamefont {Ventriglia}}]{Ibort:2009}%
  \BibitemOpen
  \bibfield  {author} {\bibinfo {author} {\bibfnamefont {A.}~\bibnamefont {Ibort}}, \bibinfo {author} {\bibfnamefont {V.~I.}\ \bibnamefont {Man{\textquotesingle}ko}}, \bibinfo {author} {\bibfnamefont {G.}~\bibnamefont {Marmo}}, \bibinfo {author} {\bibfnamefont {A.}~\bibnamefont {Simoni}},\ and\ \bibinfo {author} {\bibfnamefont {F.}~\bibnamefont {Ventriglia}},\ }\bibfield  {title} {\bibinfo {title} {An introduction to the tomographic picture of quantum mechanics},\ }\href {https://doi.org/10.1088/0031-8949/79/06/065013} {\bibfield  {journal} {\bibinfo  {journal} {Physica Scripta}\ }\textbf {\bibinfo {volume} {79}},\ \bibinfo {pages} {065013} (\bibinfo {year} {2009})}\BibitemShut {NoStop}%
\bibitem [{\citenamefont {Lvovsky}\ and\ \citenamefont {Raymer}(2009)}]{Lvovsky:2009}%
  \BibitemOpen
  \bibfield  {author} {\bibinfo {author} {\bibfnamefont {A.~I.}\ \bibnamefont {Lvovsky}}\ and\ \bibinfo {author} {\bibfnamefont {M.~G.}\ \bibnamefont {Raymer}},\ }\bibfield  {title} {\bibinfo {title} {Continuous-variable optical quantum-state tomography},\ }\href {https://doi.org/10.1103/RevModPhys.81.299} {\bibfield  {journal} {\bibinfo  {journal} {Rev. Mod. Phys.}\ }\textbf {\bibinfo {volume} {81}},\ \bibinfo {pages} {299} (\bibinfo {year} {2009})}\BibitemShut {NoStop}%
\bibitem [{\citenamefont {Filippov}\ and\ \citenamefont {Man{\textquotesingle}ko}(2011)}]{Filippov:2011}%
  \BibitemOpen
  \bibfield  {author} {\bibinfo {author} {\bibfnamefont {S.~N.}\ \bibnamefont {Filippov}}\ and\ \bibinfo {author} {\bibfnamefont {V.~I.}\ \bibnamefont {Man{\textquotesingle}ko}},\ }\bibfield  {title} {\bibinfo {title} {Optical tomography of {F}ock state superpositions},\ }\href {https://doi.org/10.1088/0031-8949/83/05/058101} {\bibfield  {journal} {\bibinfo  {journal} {Physica Scripta}\ }\textbf {\bibinfo {volume} {83}},\ \bibinfo {pages} {058101} (\bibinfo {year} {2011})}\BibitemShut {NoStop}%
\bibitem [{\citenamefont {Fisher}\ \emph {et~al.}(1984)\citenamefont {Fisher}, \citenamefont {Nieto},\ and\ \citenamefont {Sandberg}}]{Fisher:1984}%
  \BibitemOpen
  \bibfield  {author} {\bibinfo {author} {\bibfnamefont {R.~A.}\ \bibnamefont {Fisher}}, \bibinfo {author} {\bibfnamefont {M.~M.}\ \bibnamefont {Nieto}},\ and\ \bibinfo {author} {\bibfnamefont {V.~D.}\ \bibnamefont {Sandberg}},\ }\bibfield  {title} {\bibinfo {title} {Impossibility of naively generalizing squeezed coherent states},\ }\href {https://doi.org/10.1103/PhysRevD.29.1107} {\bibfield  {journal} {\bibinfo  {journal} {Phys. Rev. D}\ }\textbf {\bibinfo {volume} {29}},\ \bibinfo {pages} {1107} (\bibinfo {year} {1984})}\BibitemShut {NoStop}%
\bibitem [{\citenamefont {Gottesman}\ \emph {et~al.}(2001)\citenamefont {Gottesman}, \citenamefont {Kitaev},\ and\ \citenamefont {Preskill}}]{Preskill:2001}%
  \BibitemOpen
  \bibfield  {author} {\bibinfo {author} {\bibfnamefont {D.}~\bibnamefont {Gottesman}}, \bibinfo {author} {\bibfnamefont {A.}~\bibnamefont {Kitaev}},\ and\ \bibinfo {author} {\bibfnamefont {J.}~\bibnamefont {Preskill}},\ }\bibfield  {title} {\bibinfo {title} {Encoding a qubit in an oscillator},\ }\href {https://doi.org/10.1103/PhysRevA.64.012310} {\bibfield  {journal} {\bibinfo  {journal} {Phys. Rev. A}\ }\textbf {\bibinfo {volume} {64}},\ \bibinfo {pages} {012310} (\bibinfo {year} {2001})}\BibitemShut {NoStop}%
\bibitem [{\citenamefont {Arman}\ and\ \citenamefont {Panigrahi}(2024)}]{Arman:2024}%
  \BibitemOpen
  \bibfield  {author} {\bibinfo {author} {\bibnamefont {Arman}}\ and\ \bibinfo {author} {\bibfnamefont {P.~K.}\ \bibnamefont {Panigrahi}},\ }\bibfield  {title} {\bibinfo {title} {Generating overlap between compass states and squeezed, displaced, or {F}ock states},\ }\href {https://doi.org/10.1103/PhysRevA.109.033724} {\bibfield  {journal} {\bibinfo  {journal} {Phys. Rev. A}\ }\textbf {\bibinfo {volume} {109}},\ \bibinfo {pages} {033724} (\bibinfo {year} {2024})}\BibitemShut {NoStop}%
\end{thebibliography}%


%apsrev4-2.bst 2019-01-14 (MD) hand-edited version of apsrev4-1.bst
%Control: key (0)
%Control: author (8) initials jnrlst
%Control: editor formatted (1) identically to author
%Control: production of article title (0) allowed
%Control: page (0) single
%Control: year (1) truncated
%Control: production of eprint (0) enabled
\begin{thebibliography}{2}%
\makeatletter
\providecommand \@ifxundefined [1]{%
 \@ifx{#1\undefined}
}%
\providecommand \@ifnum [1]{%
 \ifnum #1\expandafter \@firstoftwo
 \else \expandafter \@secondoftwo
 \fi
}%
\providecommand \@ifx [1]{%
 \ifx #1\expandafter \@firstoftwo
 \else \expandafter \@secondoftwo
 \fi
}%
\providecommand \natexlab [1]{#1}%
\providecommand \enquote  [1]{``#1''}%
\providecommand \bibnamefont  [1]{#1}%
\providecommand \bibfnamefont [1]{#1}%
\providecommand \citenamefont [1]{#1}%
\providecommand \href@noop [0]{\@secondoftwo}%
\providecommand \href [0]{\begingroup \@sanitize@url \@href}%
\providecommand \@href[1]{\@@startlink{#1}\@@href}%
\providecommand \@@href[1]{\endgroup#1\@@endlink}%
\providecommand \@sanitize@url [0]{\catcode `\\12\catcode `\$12\catcode `\&12\catcode `\#12\catcode `\^12\catcode `\_12\catcode `\%12\relax}%
\providecommand \@@startlink[1]{}%
\providecommand \@@endlink[0]{}%
\providecommand \url  [0]{\begingroup\@sanitize@url \@url }%
\providecommand \@url [1]{\endgroup\@href {#1}{\urlprefix }}%
\providecommand \urlprefix  [0]{URL }%
\providecommand \Eprint [0]{\href }%
\providecommand \doibase [0]{https://doi.org/}%
\providecommand \selectlanguage [0]{\@gobble}%
\providecommand \bibinfo  [0]{\@secondoftwo}%
\providecommand \bibfield  [0]{\@secondoftwo}%
\providecommand \translation [1]{[#1]}%
\providecommand \BibitemOpen [0]{}%
\providecommand \bibitemStop [0]{}%
\providecommand \bibitemNoStop [0]{.\EOS\space}%
\providecommand \EOS [0]{\spacefactor3000\relax}%
\providecommand \BibitemShut  [1]{\csname bibitem#1\endcsname}%
\let\auto@bib@innerbib\@empty
%</preamble>
\bibitem [{\citenamefont {Shanta}\ \emph {et~al.}(1994)\citenamefont {Shanta}, \citenamefont {Chaturvedi}, \citenamefont {Srinivasan}, \citenamefont {Agarwal},\ and\ \citenamefont {Mehta}}]{Shanta:1994}%
  \BibitemOpen
  \bibfield  {author} {\bibinfo {author} {\bibfnamefont {P.}~\bibnamefont {Shanta}}, \bibinfo {author} {\bibfnamefont {S.}~\bibnamefont {Chaturvedi}}, \bibinfo {author} {\bibfnamefont {V.}~\bibnamefont {Srinivasan}}, \bibinfo {author} {\bibfnamefont {G.~S.}\ \bibnamefont {Agarwal}},\ and\ \bibinfo {author} {\bibfnamefont {C.~L.}\ \bibnamefont {Mehta}},\ }\bibfield  {title} {\bibinfo {title} {Unified approach to multiphoton coherent states},\ }\href {https://doi.org/10.1103/PhysRevLett.72.1447} {\bibfield  {journal} {\bibinfo  {journal} {Phys. Rev. Lett.}\ }\textbf {\bibinfo {volume} {72}},\ \bibinfo {pages} {1447} (\bibinfo {year} {1994})}\BibitemShut {NoStop}%
\bibitem [{\citenamefont {Laha}\ \emph {et~al.}(2018)\citenamefont {Laha}, \citenamefont {Lakshmibala},\ and\ \citenamefont {Balakrishnan}}]{Laha:2018}%
  \BibitemOpen
  \bibfield  {author} {\bibinfo {author} {\bibfnamefont {P.}~\bibnamefont {Laha}}, \bibinfo {author} {\bibfnamefont {S.}~\bibnamefont {Lakshmibala}},\ and\ \bibinfo {author} {\bibfnamefont {V.}~\bibnamefont {Balakrishnan}},\ }\bibfield  {title} {\bibinfo {title} {Estimation of nonclassical properties of multiphoton coherent states from optical tomograms},\ }\href {https://doi.org/10.1080/09500340.2018.1454527} {\bibfield  {journal} {\bibinfo  {journal} {Journal of Modern Optics}\ }\textbf {\bibinfo {volume} {65}},\ \bibinfo {pages} {1466} (\bibinfo {year} {2018})}\BibitemShut {NoStop}%
\end{thebibliography}%

\end{document}

\title{Supplementary Material: Optimal sensing of photon addition and subtraction on nonclassical light}
\author{Soumyabrata Paul}
\email{soumyabrata@physics.iitm.ac.in}
\affiliation{Department of Physics, Indian Institute of Technology Madras, Chennai 600036, India}
\affiliation{Center for Quantum Information, Communication and Computing (CQuICC), Indian Institute of Technology Madras, Chennai 600036, India}
\author{Arman}
\affiliation{Department of Physical Sciences, Indian Institute of Science Education and Research Kolkata, Mohanpur 741246, India}
\author{S. Lakshmibala}
\affiliation{Center for Quantum Information, Communication and Computing (CQuICC), Indian Institute of Technology Madras, Chennai 600036, India}
\author{Prasanta K. Panigrahi}
\affiliation{Department of Physical Sciences, Indian Institute of Science Education and Research Kolkata, Mohanpur 741246, India}
\affiliation{Center for Quantum Science and Technology (CQST), Siksha `O' Anusandhan University, Khandigiri Square, Bhubaneswar 751030, India}
\author{S. Ramanan}
\affiliation{Department of Physics, Indian Institute of Technology Madras, Chennai 600036, India}
\affiliation{Center for Quantum Information, Communication and Computing (CQuICC), Indian Institute of Technology Madras, Chennai 600036, India}
\author{V. Balakrishnan}
\affiliation{Center for Quantum Information, Communication and Computing (CQuICC), Indian Institute of Technology Madras, Chennai 600036, India}

\date{\today}% It is always \today, today,
             %  but any date may be explicitly specified

%\keywords{Suggested keywords}%Use showkeys class option if keyword
                              %display desired
\maketitle

%\tableofcontents

\section{Zoomed-in tomograms corresponding to $|\xi\rangle$ and its photon added and subtracted states $|\xi,m\rangle$\label{sec:zoomed_in_tomograms_svs}} %for the photon added and subtracted squeezed vacuum state
%%%%%%%%%%%%%%%%%%%%%%%%%%%%%%%%%%%%%%%%%%%%%%%%%%%%
%%%%%%%%%%%%%%%%%%%%%%%%%%%%%%%%%%%%%%%%%%%%%%%%%%%%
\begin{figure}[h]
    \centering
    \includegraphics[width=0.45\textwidth]{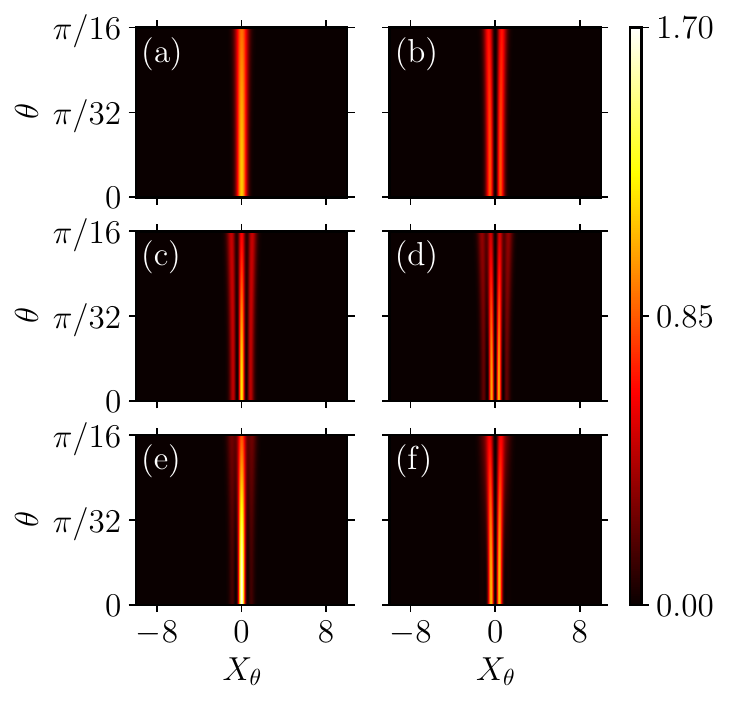}
    \caption{Left to right: Zoomed-in single-mode optical tomograms corresponding to (a) $|\xi\rangle$, (b) $|\xi,1\rangle$ (or $|\xi,-1\rangle$), (c) $|\xi,2\rangle$, (d) $|\xi,3\rangle$, (e) $|\xi,-2\rangle$, and (f) $|\xi,-3\rangle$ from $\theta = 0$ to $\theta = \pi/16$. The parameter values are as in Fig.~1 of the main text. The shift in the tomographic intensity pattern of the SVS $|\xi\rangle$ with addition or subtraction of photons shown in Fig.~1 of the main text is magnified here for better clarity. It is clear that there are striking similarities between tomograms corresponding to states which are superpositions of even Fock states (equivalently, for states which are superpositions of odd Fock states) particularly in the intensity of the central bright (equivalently dark) fringe. The difference between such states is clearly brought out in the PDFs for various $\theta$ values, as shown in Fig.~\ref{fig:fig_PDF_SingleTriplePhSubtractedSVS_theta_0} below for $|\xi,-1\rangle$ and $|\xi,-3\rangle$ for $\theta=0$.}
    \label{fig:fig_tomogram_SVS_PASVS_PSSVS_DPASVS_TPASVS_DPSSVS_TPSSVS_r_1OverSqrt2_phi_0_panel_zoomed}
\end{figure}
%%%%%%%%%%%%%%%%%%%%%%%%%%%%%%%%%%%%%%%%%%%%%%%%%%%%
%%%%%%%%%%%%%%%%%%%%%%%%%%%%%%%%%%%%%%%%%%%%%%%%%%%%

\section{Photon Added Squeezed Vacuum States\label{sec:photon_added_svs}}

%%%%%%%%%%%%%%%%%%%%%%%%%%%%%%%%%%%%%%%%%%%%%%%%%%%%
%%%%%%%%%%%%%%%%%%%%%%%%%%%%%%%%%%%%%%%%%%%%%%%%%%%%
\begin{figure}[h]
    \centering
    \includegraphics[width=0.45\textwidth]{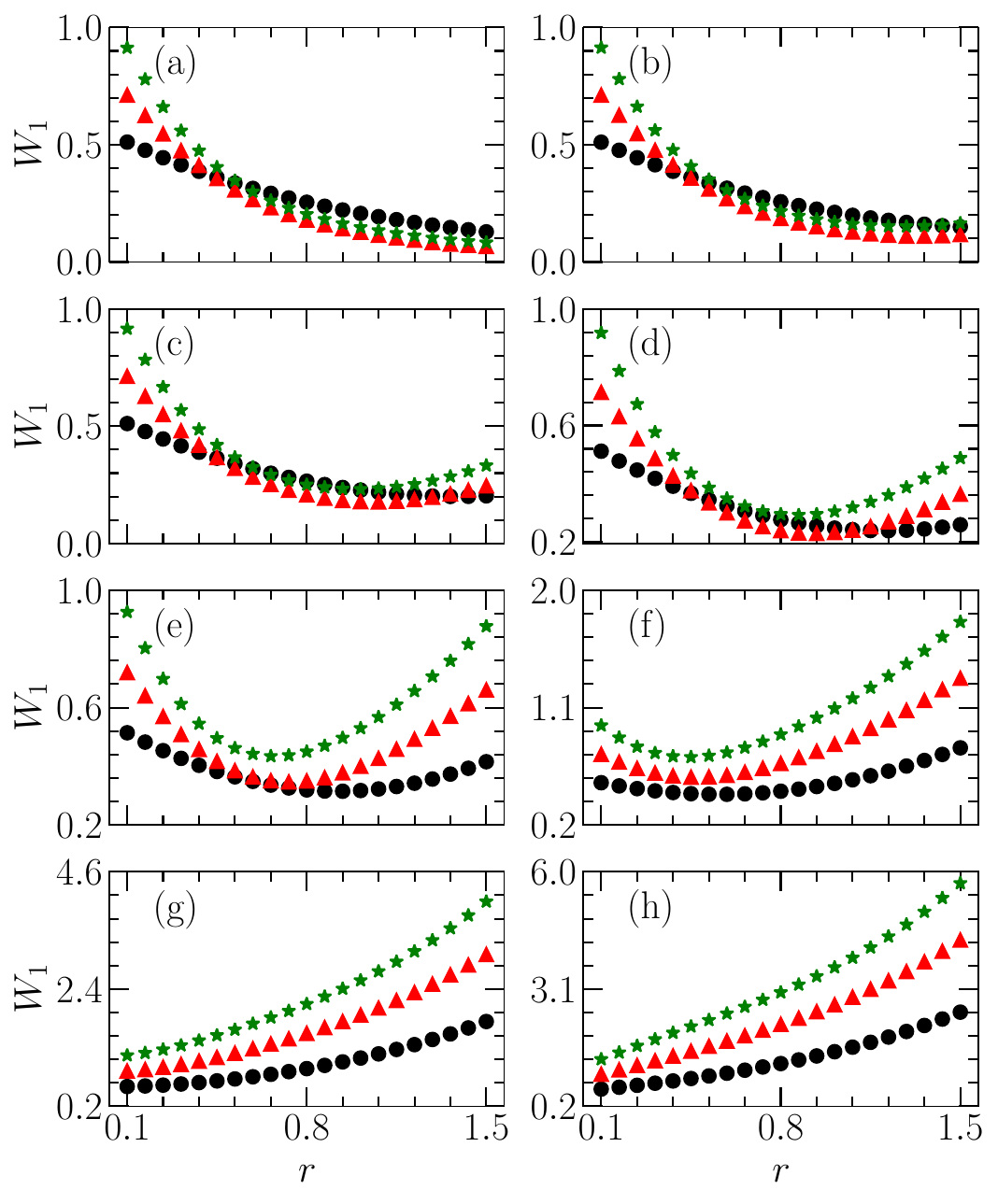}
    \caption{Plots of $W_{1}$ between $|\xi, 1\rangle$ (equivalently, $|\xi,-1\rangle$) and $|\xi\rangle$ (black circles), between $|\xi, 2\rangle$ and $|\xi\rangle$ (red triangles), and between $|\xi, 3\rangle$ and $|\xi\rangle$ (green asterisks) as functions of the squeezing parameter $r$, for $\theta =$ (a) $0$, (b) $\pi/100$, (c) $\pi/50$, (d) $\pi/35$, (e) $\pi/20$, (f) $\pi/10$, (g) $\pi/4$ and (h) $\pi/2$. The other parameters are as in Fig.~1 of the main text. We note that the value of $r$ at crossovers depends both on the quadrature and the states considered. Therefore to distinguish between two states using the Wasserstein distance, values of $r$ in the neighbourhood of the crossover must be avoided. For $\theta < \pi/20$ ($\theta > \pi/20$) crossovers are present (absent).}
    \label{fig:WD_r_theta_SVS_panel}
\end{figure}
%%%%%%%%%%%%%%%%%%%%%%%%%%%%%%%%%%%%%%%%%%%%%%%%%%%%
%%%%%%%%%%%%%%%%%%%%%%%%%%%%%%%%%%%%%%%%%%%%%%%%%%%%
We now demonstrate the occurrence of crossovers of $W_{1}$ corresponding to $|\xi\rangle$ and $|\xi,1\rangle$ (equivalently, between $|\xi\rangle$ and $|\xi,-1\rangle$), corresponding to $|\xi\rangle$ and $|\xi,2\rangle$, and corresponding to $|\xi\rangle$ and $|\xi,3\rangle$. Figs.~\ref{fig:WD_r_theta_SVS_panel} (a), (b), (c) and (d) show crossovers as a function of $r$ for $\theta = 0, \pi/100, \pi/50$ and $\pi/35$. This is a gradual transition and at $\theta=\pi/20$ (Fig.~\ref{fig:WD_r_theta_SVS_panel} (e)) the Wasserstein distances are equal at a specific value of $r$. With increase in $\theta$ crossovers are absent (Figs.~\ref{fig:WD_r_theta_SVS_panel} (f), (g) and (h)). A similar analysis can be carried out for other states and other quadratures.

%%%%%%%%%%%%%%%%%%%%%%%%%%%%%%%%%%%%%%%%%%%%%%%%%%%%
%%%%%%%%%%%%%%%%%%%%%%%%%%%%%%%%%%%%%%%%%%%%%%%%%%%%
\begin{figure}
    \centering
    \includegraphics[width=0.45\textwidth]{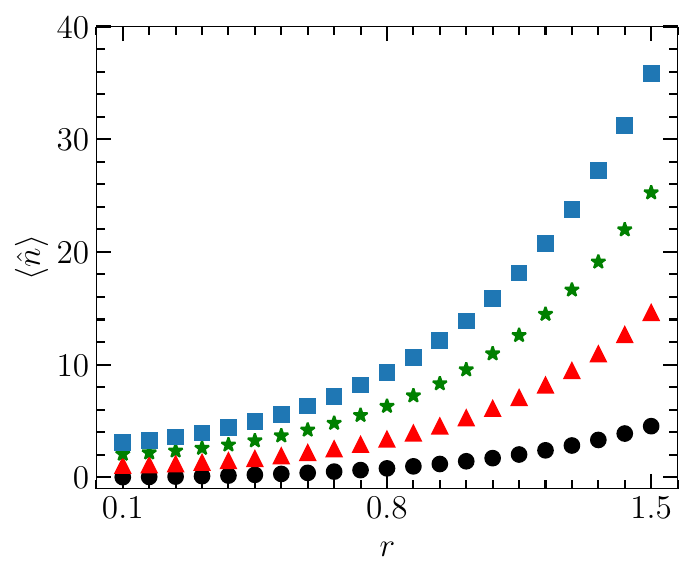}
    \caption{Plots of the mean photon number $\langle {\hat n} \rangle$, for $|\xi\rangle$ (black circles), $|\xi,1\rangle$ (red triangles), $|\xi,2\rangle$ (green asterisks), and $|\xi,3\rangle$ (blue squares) as functions of the squeezing parameter $r$.
    For any given $r$ we see that photon addition to $|\xi\rangle$ increases $\langle {\hat n} \rangle$, and hence no two states have the same $\langle {\hat n} \rangle$.}
    \label{fig:mean_photon_num_0_1_2_3_PhAddedSVS}
\end{figure}
%%%%%%%%%%%%%%%%%%%%%%%%%%%%%%%%%%%%%%%%%%%%%%%%%%%%
%%%%%%%%%%%%%%%%%%%%%%%%%%%%%%%%%%%%%%%%%%%%%%%%%%%%

%%%%%%%%%%%%%%%%%%%%%%%%%%%%%%%%%%%%%%%%%%%%%%%%%%%%
%%%%%%%%%%%%%%%%%%%%%%%%%%%%%%%%%%%%%%%%%%%%%%%%%%%%
\begin{figure}
    \centering
    \includegraphics[width=0.45\textwidth]{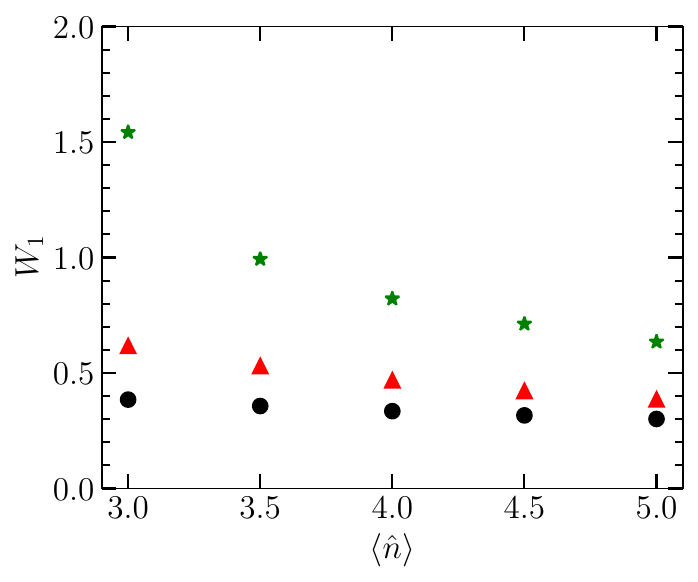}
    \caption{Plots of $W_{1}$ between $|\xi, 1\rangle$ (equivalently, $|\xi,-1\rangle$) and $|\xi\rangle$ (black circles), between $|\xi, 2\rangle$ and $|\xi\rangle$ (red triangles), and between $|\xi, 3\rangle$ and $|\xi\rangle$ (green asterisks) as functions of $\langle {\hat n} \rangle$, along the $x$-quadrature. The other parameters are as in Fig.~1 of the main text. Increasing $\langle {\hat n} \rangle$ results in an increase in $r$ (Fig.~3 above), and $W_{1}$ is a decreasing function of $r$ (Fig.~3 of the main text). Hence $W_{1}$ is a decreasing function of $\langle {\hat n} \rangle$ for all the three states. Furthermore with increase in the number of added photons, $W_{1}$ increases, indicating no crossover between states with the same $\langle {\hat n} \rangle$.}
    \label{fig:wass_dist_mean_photon_number_equal_PhAddedSVS}
\end{figure}
%%%%%%%%%%%%%%%%%%%%%%%%%%%%%%%%%%%%%%%%%%%%%%%%%%%%
%%%%%%%%%%%%%%%%%%%%%%%%%%%%%%%%%%%%%%%%%%%%%%%%%%%%

We have considered the manner in which $W_{1}$ for the SVS and its $1$, $2$, and $3$ photon added counterparts varies with the mean photon number $\langle {\hat n} \rangle$. This is illustrated through Figs.~\ref{fig:mean_photon_num_0_1_2_3_PhAddedSVS} and~\ref{fig:wass_dist_mean_photon_number_equal_PhAddedSVS}. From Fig.~\ref{fig:mean_photon_num_0_1_2_3_PhAddedSVS} we see that for a given value of the squeezing parameter $r$, photon addition to $|\xi\rangle$ increases $\langle {\hat n} \rangle$, and hence no two states have the same $\langle {\hat n} \rangle$. (We also note that as $r \rightarrow 0$, $\langle {\hat n} \rangle_{|\xi, m\rangle} \rightarrow m$, the number of added photons to $|\xi\rangle$.) This feature taken together with the fact that $W_{1}$ is a decreasing function of $r$ (Fig.~3 of the main text) implies that $W_{1}$ is a decreasing function of $\langle {\hat n} \rangle$ as well for all the three states. Furthermore with increase in the number of added photons, $W_{1}$ increases, indicating no crossover between states with the same photon number $\langle {\hat n} \rangle$. This is shown explicitly in Fig.~\ref{fig:wass_dist_mean_photon_number_equal_PhAddedSVS}. A similar exercise can be carried out for the photon subtracted states as well.

\begin{figure}[h]
    \centering
    \includegraphics[width=0.45\textwidth]{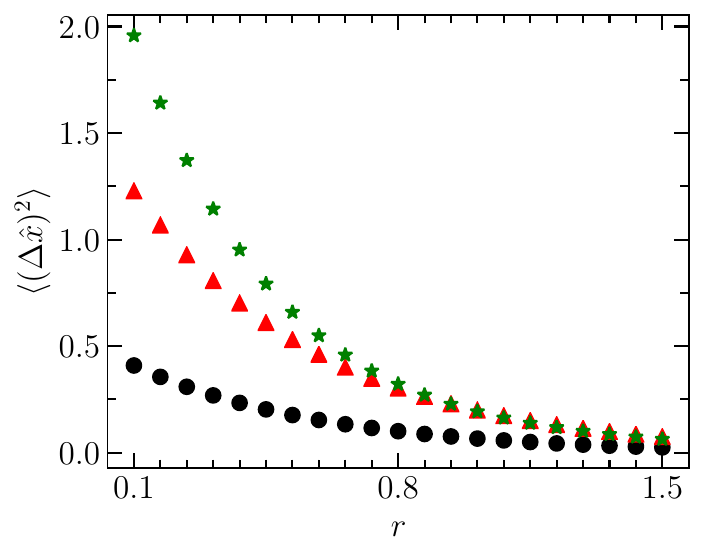}
    \caption{The variance $\langle (\Delta {\hat x})^{2} \rangle$ as a function of the squeezing parameter $r$ for $|\xi\rangle$ (black circles), $|\xi,1\rangle$ (red triangles), and $|\xi,2\rangle$ (green asterisks). The other parameters are as in Fig.~1 of the main text. $\langle (\Delta {\hat x})^{2} \rangle \sim \exp(-\kappa r)$. Smaller values of $\kappa$ correspond to more addition of photons.}
    \label{fig:Deltax_r_PhAddedSVS}
\end{figure}

It is also interesting to consider the quadrature variance $\langle (\Delta {\hat x})^{2} \rangle$ as a function of $r$ (Fig.~\ref{fig:Deltax_r_PhAddedSVS}), since it is easily calculable from the tomogram. As expected, for  all the three states ($|\xi\rangle$, $|\xi,1\rangle$ and $|\xi,2\rangle$),  $\langle (\Delta {\hat x})^{2} \rangle \sim \exp(-\kappa r)$, with $\kappa = 2$ for $|\xi\rangle$ and $\kappa < 2$ with addition of photons. The smaller value of $\kappa$ corresponds to more photon addition.
%%%%%%%%%%%%%%%%%%%%%%%%%%%%%%%%%%%%%%%%%%%%%%%%%%%%
%%%%%%%%%%%%%%%%%%%%%%%%%%%%%%%%%%%%%%%%%%%%%%%%%%%%
%%%%%%%%%%%%%%%%%%%%%%%%%%%%%%%%%%%%%%%%%%%%%%%%%%%%
%%%%%%%%%%%%%%%%%%%%%%%%%%%%%%%%%%%%%%%%%%%%%%%%%%%%

\begin{table}[b]
\caption{\label{tab:expansion_coeffs}
We tabulate below the different states used in the main text and their relevant probability amplitudes.
Here, $\kappa(n)=(-e^{i\phi}\tanh{r})^{n}/(2^{n}n!)$.}
\begin{ruledtabular}
\begin{tabular}{lr}
$\ket{\psi}=\sum c_{m}|m\rangle$&
$c_{m}$
\\
\colrule
$\ket{\xi}$& $\delta_{m,2n}\frac{\kappa(n)\sqrt{(2n)!}}{\sqrt{\cosh r}}$ \\
$\ket{\xi,\pm1}$& $\delta_{m,2n+1}\frac{\kappa(n)\sqrt{(2n+1)!}}{\sqrt{(\cosh^{3}r)}}$ \\
$\ket{\xi,2}$& $\delta_{m,2n+2}\frac{\kappa(n)\sqrt{(2n+2)}!}{\sqrt{\left( 3\cosh^{5}r - \cosh^{3}r \right)}}$ \\
$\ket{\xi,3}$& $\delta_{m,2n+3}\frac{\kappa(n)\sqrt{(2n+3)}!}{\sqrt{\left( 15\cosh^{7}r - 9\cosh^{5}r \right)}}$ \\
$\ket{\xi,-2}$& $\delta_{m,2n-2}\frac{\kappa(n)(2n)!}{\sqrt{(2n-2)!}\sqrt{(3\cosh^{5}r - 5\cosh^{3}r+2\cosh r)}}$ \\ 
$\ket{\xi,-3}$& $\delta_{m,2n-3}\frac{\kappa(n)(2n)!}{\sqrt{(2n-3)!}\sqrt{3\sinh^{4}r(5\cosh^{3}r-2\cosh{r})}}$ \\ 
%$\ket{\xi,-1}$& $\delta_{m,2n-1}\frac{\kappa(n)\,(2n)!}{\sqrt{(2n-1)!}\sqrt{(\cosh r)^{3}}}$ \\

\end{tabular}
\end{ruledtabular}
\end{table}

\section{Photon Subtracted Squeezed Vacuum States\label{sec:photon_subtracted_svs}}

\begin{figure}[h]
    \centering
    \includegraphics[width=0.45\textwidth]{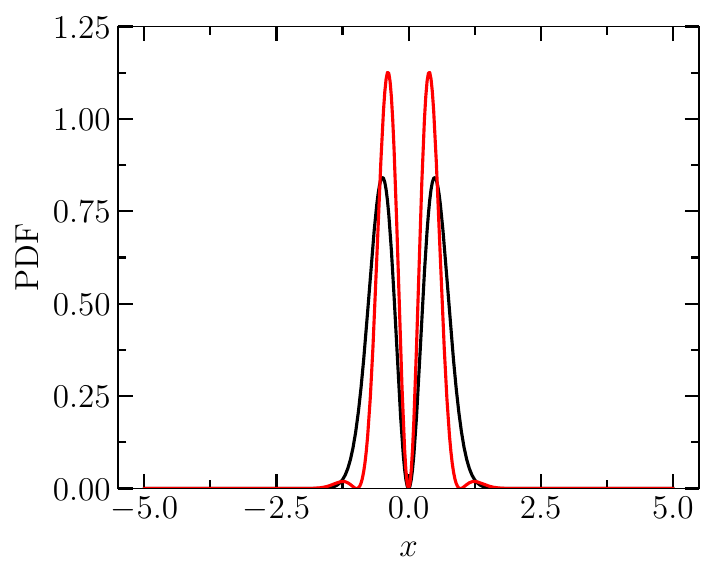}
    \caption{Probability distribution functions (PDFs) along the $x$-quadrature corresponding to $|\xi,-1\rangle$ [or $|\xi,1\rangle$] (black) and $|\xi,-3\rangle$ (red) for $r = 1/\sqrt{2}$. The other parameter values are as in Fig.~1 of the main text. The differences between other pairs of states can be clearly seen by comparing their corresponding PDFs in specific quadratures.}
    %This corroborates the result inferred from Fig.~\ref{fig:WD_r_theta_SVSSubtracted_panel}.}
    \label{fig:fig_PDF_SingleTriplePhSubtractedSVS_theta_0}
\end{figure}

The PDFs for $|\xi,-1\rangle$ and $|\xi,-3\rangle$ along the $x$-quadrature are shown in Fig.~\ref{fig:fig_PDF_SingleTriplePhSubtractedSVS_theta_0}.
The differences between the two PDFs, though discernible, are small in magnitude because all states which arise by removing an odd number of photons
from the SVS, still have a contribution from $|0\rangle$.
This corroborates the similarities between the corresponding tomograms in Fig.~1 of the main text.

\begin{figure}
    \centering
    \includegraphics[width=0.45\textwidth]{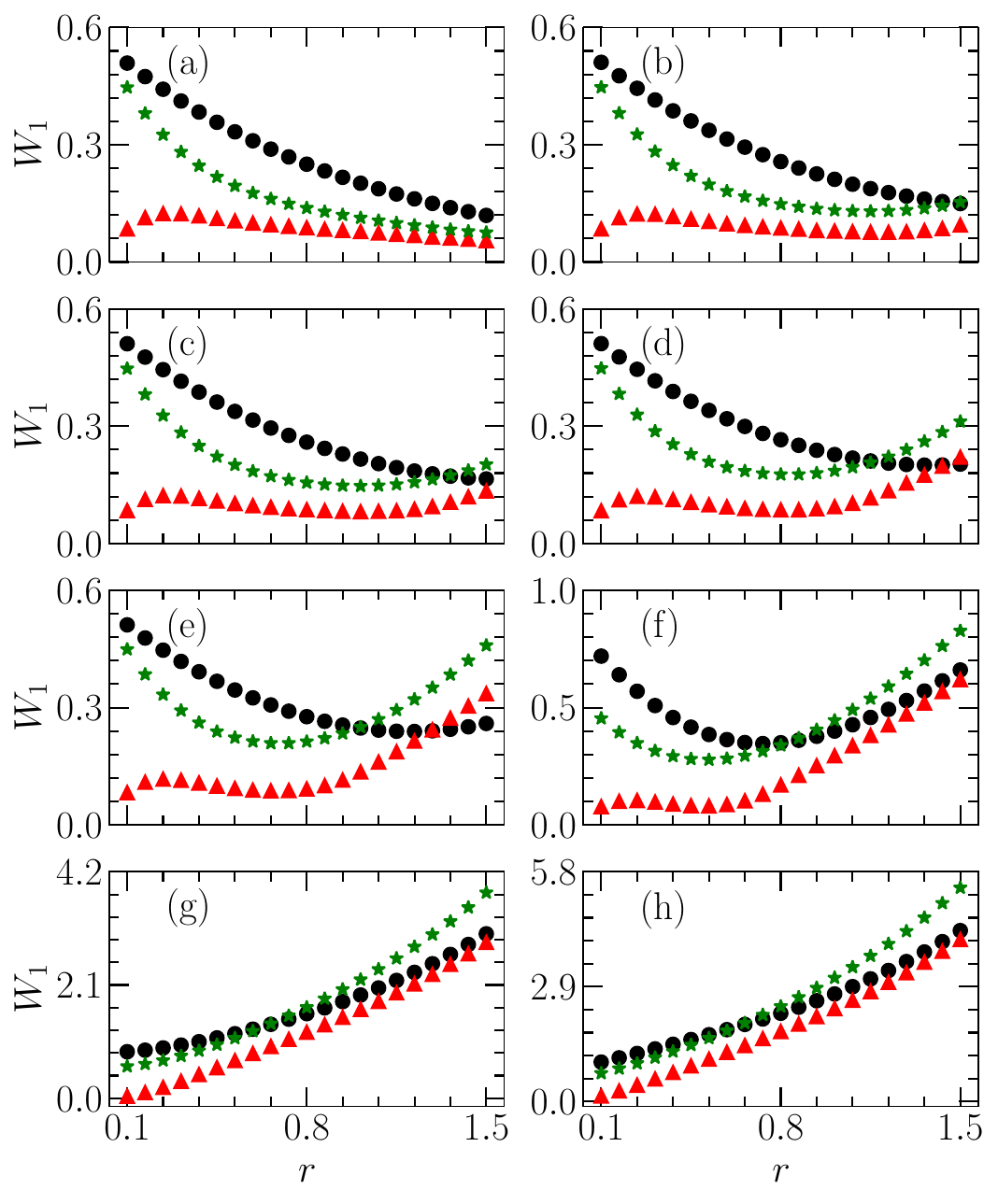}
    \caption{Plots of the Wasserstein distance $W_{1}$ between $|\xi, -1\rangle$ (equivalently, $|\xi,1\rangle$) and $|\xi\rangle$ (black circles), between $|\xi, -2\rangle$ and $|\xi\rangle$ (red triangles), and between $|\xi, -3\rangle$ and $|\xi\rangle$ (green asterisks) as a function of the squeezing parameter $r$, for $\theta =$ (a) $0$, (b) $\pi/100$, (c) $\pi/75$, (d) $\pi/50$, (e) $\pi/35$, (f) $\pi/20$, (g) $\pi/4$ and (h) $\pi/2$. The other parameters are as in Fig.~1 of the main text. We note that the value of $r$ at crossovers depends both on the quadrature and the states considered. In particular, in contrast to photon addition (see Fig.~\ref{fig:WD_r_theta_SVS_panel} above) there is no crossover for $\theta=0$ in the range of $r$ considered.}
    \label{fig:WD_r_theta_SVSSubtracted_panel}
\end{figure}

In Fig.~\ref{fig:WD_r_theta_SVSSubtracted_panel} plots of the Wasserstein distances (between $|\xi\rangle$ and $|\xi,-1\rangle$, between $|\xi\rangle$ and $|\xi,-2\rangle$, and between $|\xi\rangle$ and $|\xi,-3\rangle$ functions of $r$ have been shown.
%%%%%%%%%%%%%%%%%%%%%%%%%%%%%%%%%%%%%%%%%%%%%%%%%%%%
%%%%%%%%%%%%%%%%%%%%%%%%%%%%%%%%%%%%%%%%%%%%%%%%%%%%
%%%%%%%%%%%%%%%%%%%%%%%%%%%%%%%%%%%%%%%%%%%%%%%%%%%%
%%%%%%%%%%%%%%%%%%%%%%%%%%%%%%%%%%%%%%%%%%%%%%%%%%%%

%%%%%%%%%%%%%%%%%%%%%%%%%%%%%%%%%%%%%%%%%%%%%%%%%%%%
%%%%%%%%%%%%%%%%%%%%%%%%%%%%%%%%%%%%%%%%%%%%%%%%%%%%
%%%%%%%%%%%%%%%%%%%%%%%%%%%%%%%%%%%%%%%%%%%%%%%%%%%%
%%%%%%%%%%%%%%%%%%%%%%%%%%%%%%%%%%%%%%%%%%%%%%%%%%%%

\section{Photon Added Even Coherent States\label{sec:photon_added_ecs}}

We now proceed to examine the effect of photon addition to the ECS. The ECS is referred to, as the `Janus-faced' partner of the SVS because of the following reason~\cite{Shanta:1994}. 
Since both are superpositions of even Fock states, the photon creation and destruction operators of direct relevance are $\hat{a}^{\dagger 2}$ and $\hat{a}^{2}$. Using $[\hat{a}, \hat{a}^{\dagger}] = 1$, it can be easily verified that 
\begin{equation}
[\hat{a}^{2}, \hat{G}_{0}^{\dagger}] = 1,
\label{eq:janus_faced_ops_commutator}
\end{equation}
with
\begin{equation}
\hat{G}_{0}^{\dagger} = \hat{a}^{\dagger 2} (1 + \hat{a}^{\dagger} \hat{a})^{-1}.
\label{eq:janus_faced_ops_definition}
\end{equation}
The normalized ECS $|\alpha\rangle_{+} = \left(|\alpha\rangle + |-\alpha\rangle\right)/
\{2(1+\exp(-2|\alpha|^{2}))\}^{1/2}$ (where $|\alpha\rangle$ is the coherent state, and $\alpha \in \mathbb{C}$),
can be expressed as $\exp(f\hat{G}_{0}^{\dagger})|0\rangle$ with the identification $\alpha = \sqrt{f}$. As indicated in Sec.~III of the main text, the SVS is obtained by the action of the exponential of $\hat{a}^{\dagger 2}$ on the vacuum.
This feature together with Eq.~\eqref{eq:janus_faced_ops_commutator} reveals an interesting connection between the ECS and the SVS. It also suggests there could be similarities in the tomographic aspects of the ECS and the SVS with the addition of photons.

%%%%%%%%%%%%%%%%%%%%%%%%%%%%%%%%%%%%%%%%%%%%%%%%%%%%
%%%%%%%%%%%%%%%%%%%%%%%%%%%%%%%%%%%%%%%%%%%%%%%%%%%%
\begin{figure}[h]
    \centering
    \includegraphics[width=0.45\textwidth]{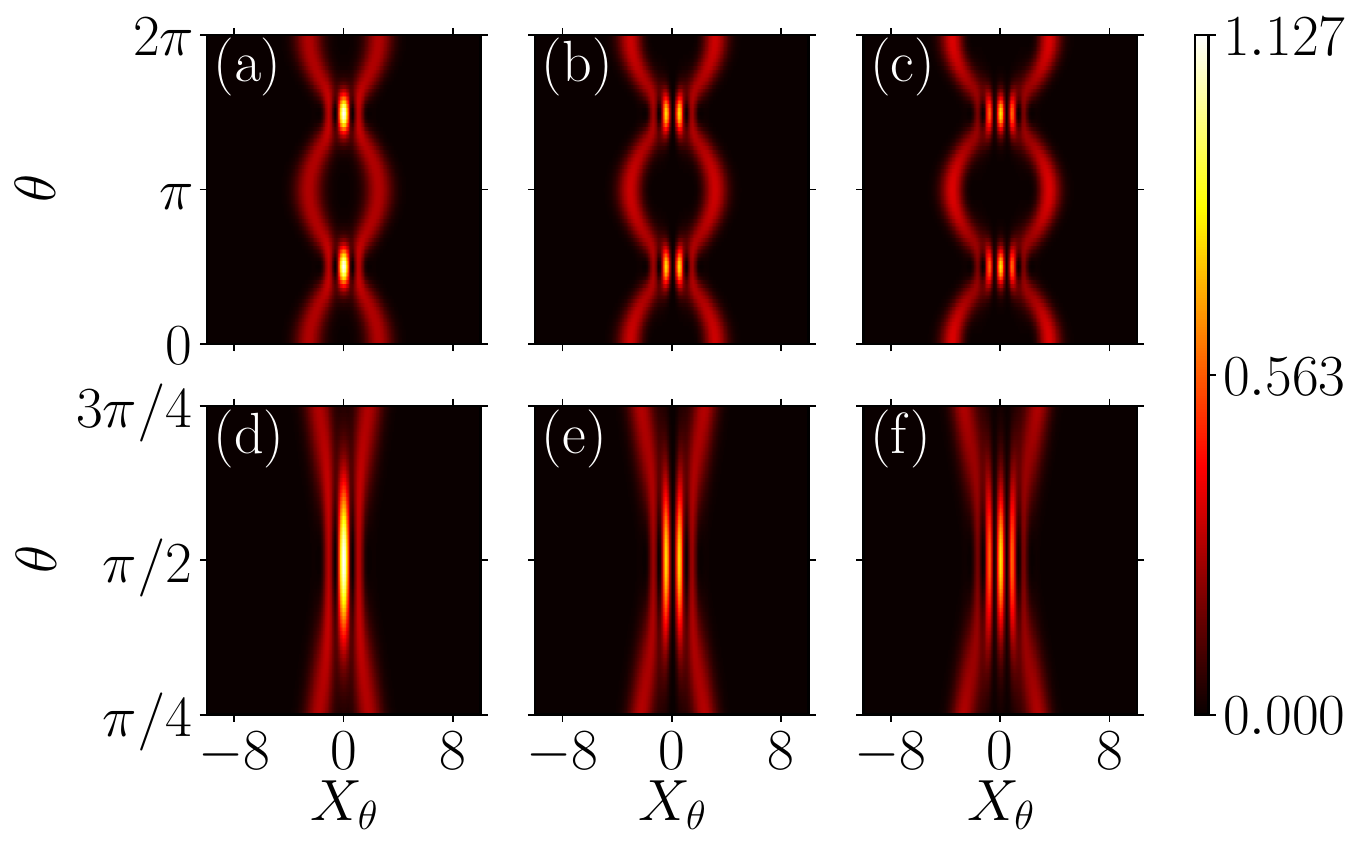}
    \caption{Top panel: Left to right: Single-mode optical tomograms for the ECS and its photon added counterparts respectively, namely (a) $|\alpha\rangle_{+}$, (b) $|\alpha_{+}, 1\rangle$, and (c) $|\alpha_{+}, 2\rangle$. We set $\alpha = 1.8$, so that $|\alpha\rangle$ and $|-\alpha\rangle$ (comprising the ECS) are sufficiently apart, and for the interference fringes are clearly visible. Notice that as in the case of the SVS, intensity shifts are seen at $\theta=\pi/2$ and $3\pi/2$ with addition of photons. Bottom panel: Left to right: Zoomed in version of (a) -- (c) from $\theta=\pi/4$ to $3\pi/4$.}
    \label{fig:fig_tomogram_ECS_alpha_1.8_panel}
\end{figure}
%%%%%%%%%%%%%%%%%%%%%%%%%%%%%%%%%%%%%%%%%%%%%%%%%%%%
%%%%%%%%%%%%%%%%%%%%%%%%%%%%%%%%%%%%%%%%%%%%%%%%%%%%
Analogous to the procedure carried out in Sec.~III of the main text, we now examine the effect of one and two photon addition to the ECS. In evident notation, these states can be expanded in the Fock basis as
\begin{equation}
|\alpha_{+}, 1\rangle = \mathcal{M}_{1}\sum_{n=0}^{\infty} \sqrt{\frac{2n+1}{(2n)!}} \alpha^{2n} |2n+1\rangle,
\label{eq:psi_one_photon_added_ecs_fock_expansion}
\end{equation}
where $\mathcal{M}_{1} = \left\{ \cosh|\alpha|^{2}+|\alpha|^{2}\sinh|\alpha|^{2} \right\}^{-1/2}$, and
\begin{equation}
|\alpha_{+}, 2\rangle = \mathcal{M}_{2}\sum_{n=0}^{\infty} \sqrt{\frac{(2n+2)(2n+1)}{(2n)!}} \alpha^{2n} |2n+2\rangle,
\label{eq:psi_two_photon_added_ecs_fock_expansion}
\end{equation}
with $\mathcal{M}_{2} = \left\{ (2+|\alpha|^{4})\cosh|\alpha|^{2}+4|\alpha|^{2}\sinh|\alpha|^{2} \right\}^{-1/2}$.

%%%%%%%%%%%%%%%%%%%%%%%%%%%%%%%%%%%%%%%%%%%%%%%%%%%%
%%%%%%%%%%%%%%%%%%%%%%%%%%%%%%%%%%%%%%%%%%%%%%%%%%%%
\begin{figure}[h]
    \centering
    \includegraphics[width=0.45\textwidth]{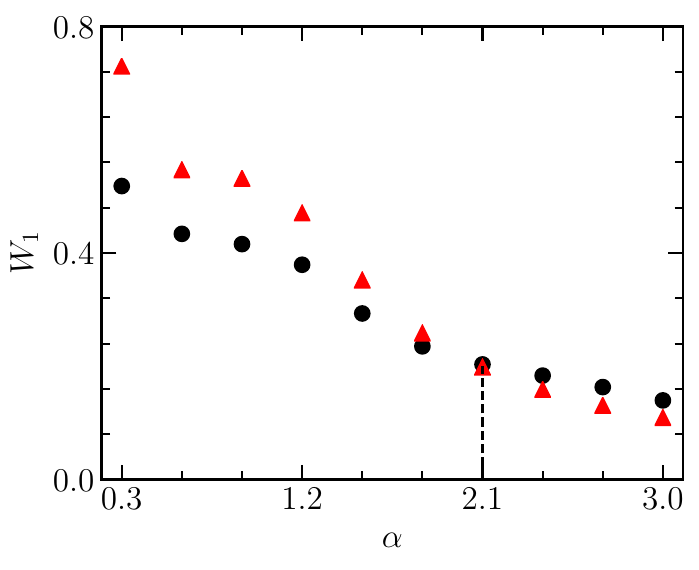}
    \caption{Wasserstein distance $W_{1}$ between $|\alpha\rangle_{+}$ and $|\alpha_{+}, 1\rangle$ (black circles), and between $|\alpha\rangle_{+}$ and $|\alpha_{+}, 2\rangle$ (red triangles), as a functions of $\alpha$. $W_{1}$ is computed by comparing the corresponding $p$-quadrature ($\theta=\pi/2$) probability distribution functions. The crossover is at $\alpha \approx 2.1$.}
    \label{fig:WD_alpha_PhAddedECS}
\end{figure}
%%%%%%%%%%%%%%%%%%%%%%%%%%%%%%%%%%%%%%%%%%%%%%%%%%%%
%%%%%%%%%%%%%%%%%%%%%%%%%%%%%%%%%%%%%%%%%%%%%%%%%%%%
The tomograms corresponding to $|\alpha\rangle_{+}$, $|\alpha_{+}, 1\rangle$ and $|\alpha_{+}, 2\rangle$, with $\alpha = 1.8$, are shown in Fig.~\ref{fig:fig_tomogram_ECS_alpha_1.8_panel}. (In each of these figures, the appearance of the two strands can be traced back to the superposition of two coherent states to form the ECS.) The interference fringes in this case are shifted by $\pi/2$ compared to the tomograms in Fig.~1 of the main text~\cite{Laha:2018}. However, the intensity shifts in these fringes on addition of photons is similar to that of the SVS. It is evident that $\theta=\pi/2$ (equiv. $3\pi/2$) are the relevant quadratures for computing $W_{1}$ between the ECS and the $1$-photon added ECS, and also between the ECS and its $2$-photon added counterpart. The two plots of $W_{1}$ vs. $\alpha$ (taken to be real) in Fig.~\ref{fig:WD_alpha_PhAddedECS} display the crossover at $\alpha \approx 2.1$, with trends similar to that seen in Fig.~2 of the main text. These trends can be explained from the individual PDFs as was done in Sec.~III of the main text.

\begin{figure}[h]
    \centering
    \includegraphics[width=0.45\textwidth]{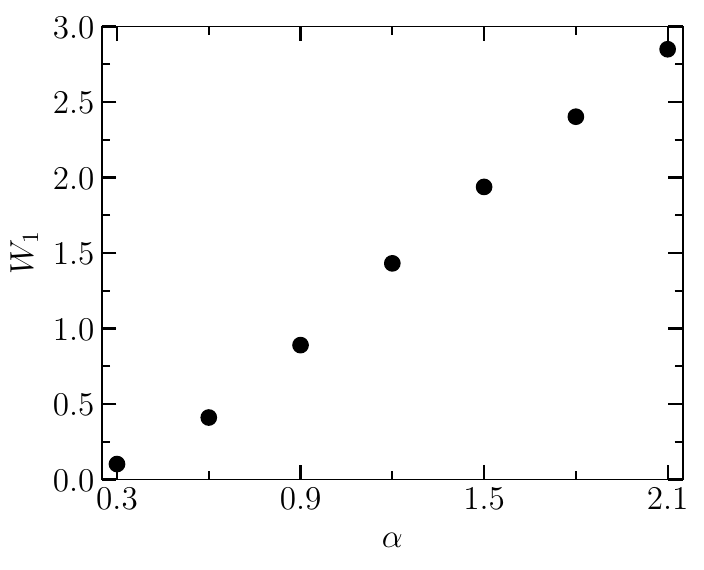}
    \caption{Wasserstein distance $W_{1}$ between $|\xi\rangle$ and $|\alpha\rangle_{+}$ as a function of the squeezing parameter $r$ (equiv. $\alpha$), along the $x$-quadrature. The mean photon number for the two states have been set to be equal.}
    \label{fig:WD_SVSandECS_theta_0}
\end{figure}

Given the Janus-faced relationship between the SVS and the ECS, it is interesting to examine the manner in which the Wasserstein distance between them varies as a function of the squeezing parameter $r$ (equiv. $\alpha$), setting $\theta = 0$ (see Fig.~\ref{fig:WD_SVSandECS_theta_0}).
Here, we have chosen $\alpha$ and $r$, such that the mean photon numbers for both the states are equal, i.e., we set $\sinh^{2}r = |\alpha|^{2}\tanh|\alpha|^{2}$.
As $r$ increases (equiv. $\alpha$ increases), $|\xi\rangle$ gets concentrated along $x=0$, whereas $|\alpha\rangle_{+}$ spreads. Consequently, as $r$ (and $\alpha$) increases, $W_{1}$ also increases. This behavior of the Wasserstein distance can be explained from the corresponding PDFs of the two states along the $x$-quadrature.

%%%%%%%%%%%%%%%%%%%%%%%%%%%%%%%%%%%%%%%%%%%%%%%%%%%%
%%%%%%%%%%%%%%%%%%%%%%%%%%%%%%%%%%%%%%%%%%%%%%%%%%%%
%%%%%%%%%%%%%%%%%%%%%%%%%%%%%%%%%%%%%%%%%%%%%%%%%%%%
%%%%%%%%%%%%%%%%%%%%%%%%%%%%%%%%%%%%%%%%%%%%%%%%%%%%

%%%%%%%%%%%%%%%%%%%%%%%%%%%%%%%%%%%%%%%%%%%%%%%%%%%%
%%%%%%%%%%%%%%%%%%%%%%%%%%%%%%%%%%%%%%%%%%%%%%%%%%%%
\begin{figure}[h]
    \centering
    \includegraphics[width=0.45\textwidth]{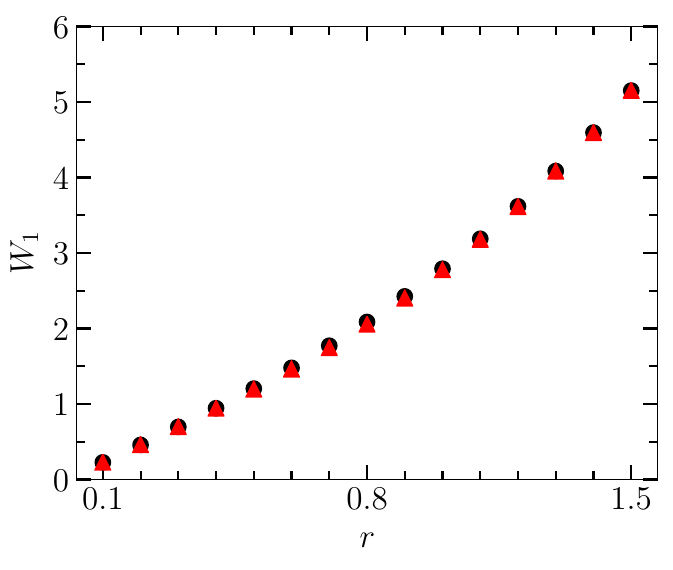}
    \caption{Wasserstein distance $W_{1}$ between (a) $|\xi,1\rangle$ and $|\alpha\rangle_{-}$ (black circles), and (b) $|\xi,1\rangle$ and $|\alpha_{+},1\rangle$ (red triangles), as a function of $r$, along the $x$-quadrature. The mean photon number for the two states in both (a) and (b) have been set to be equal.}
    \label{fig:WD_SPASVSandOCS_SPASVSandSPAECS_theta_0_vary_r.pdf}
\end{figure}
%%%%%%%%%%%%%%%%%%%%%%%%%%%%%%%%%%%%%%%%%%%%%%%%%%%%
%%%%%%%%%%%%%%%%%%%%%%%%%%%%%%%%%%%%%%%%%%%%%%%%%%%%

A similar exercise can be carried out in the case of the $1$-photon added SVS $|\xi,1\rangle$ (equiv. $|\xi,-1\rangle$) and the $1$-photon added ECS $|\alpha_{+},1\rangle$ (black circles in Fig.~\ref{fig:WD_SPASVSandOCS_SPASVSandSPAECS_theta_0_vary_r.pdf}), and $1$-photon added SVS $|\xi,1\rangle$ (equiv. $|\xi,-1\rangle$) and the odd coherent state (OCS) $|\alpha\rangle_{-}$ (red triangles in Fig.~\ref{fig:WD_SPASVSandOCS_SPASVSandSPAECS_theta_0_vary_r.pdf}). The  mean photon numbers of the two states, in each case, are set to be equal.
%Increasing $\alpha$ implies increasing $r$. 
The two lobes (corres. to $|\alpha\rangle$ and $|-\alpha\rangle$) of the state $|\alpha_{+}, 1\rangle$ move further apart as $\alpha$ increases. On the other hand, as $r$ increases the state $|\xi, 1\rangle$ gets squeezed further. Therefore $W_{1}$ between $|\alpha_{+}, 1\rangle$ and $|\xi, 1\rangle$ increases as $r$ (equiv. $\alpha$) increases. Similar arguments hold if $|\xi, 1\rangle$ and $|\alpha\rangle_{-}$ are compared. The normalized OCS is given by $|\alpha\rangle_{-} = \left(|\alpha\rangle - |-\alpha\rangle\right)/ \{2(1-\exp(-2|\alpha|^{2}))\}^{1/2}$.

\bibliography{sm}% Produces the bibliography via BibTeX.